\documentclass[11pt,a4paper]{article}

\usepackage[colorlinks=true, linkcolor=black!50!blue, urlcolor=blue, citecolor=blue, anchorcolor=blue]{hyperref}
\usepackage[font=small,labelfont=bf,margin=0mm,labelsep=period,tableposition=top]{caption}
\usepackage[a4paper,top=3cm,bottom=2.5cm,left=2.5cm,right=2.5cm,bindingoffset=0mm]{geometry}

\usepackage{graphicx,placeins}
\usepackage{float}
\usepackage{afterpage}
\usepackage{epsfig,cite}
\usepackage{amssymb}
\usepackage{amsmath}
\usepackage{dsfont}
\usepackage{multirow}
\usepackage{url}
\usepackage{xcolor,colortbl}
\usepackage{float}
\usepackage{afterpage}
\usepackage{url}
\usepackage{hyperref}
\usepackage{booktabs}
\usepackage{placeins}

\usepackage{tikz}
\usepackage{tikz-3dplot}

\usepackage{enumitem}
\usepackage{hyperref}
\usepackage{cite}

\usepackage{pifont}

\usetikzlibrary{shapes, arrows}
\usetikzlibrary{decorations.pathreplacing}
\usetikzlibrary{positioning, calc}
\tikzstyle{fitted} = [rectangle, minimum width=5cm, minimum height=1cm, text centered, draw=black, fill=red!30]
\tikzstyle{operations} = [rectangle, rounded corners, minimum width=2cm,text centered, draw=black, fill=red!30]
\tikzstyle{roundtext} = [rectangle, rounded corners, minimum width=2cm, minimum height=0.8cm, text centered, draw=black, fill=red!30]
\tikzstyle{n3py} = [rectangle, rounded corners, minimum width=3cm, minimum height=1cm, text centered, draw=black, fill=green!30]
\tikzstyle{myarrow} = [thick,->,>=stealth]
\tikzstyle{line} =[draw, -latex']
\tikzstyle{decision} = [diamond, draw, fill=red!20, text width=7.5em, text centered,  inner sep=0pt, minimum height=2em, aspect=4]
\tikzstyle{cloud} = [draw, ellipse,fill=green!20, minimum height=2em]
\tikzstyle{inout} = [rectangle, draw, fill=green!20, text width=9.5em, text centered, rounded corners, minimum height=2em, minimum width=10em]
\tikzstyle{block}=[rectangle, draw, fill=blue!20, text width=9.5em,
                   text centered, rounded corners, minimum height=2em,
                   minimum width=10em]

\definecolor{darkgreen}{rgb}{0.0, 0.5, 0.13}

\newcommand{\be}{\begin{equation}}
\newcommand{\ee}{\end{equation}}
\newcommand{\bea}{\begin{eqnarray}}
\newcommand{\eea}{\end{eqnarray}}
\newcommand{\bi}{\begin{itemize}}
\newcommand{\ei}{\end{itemize}}
\newcommand{\ben}{\begin{enumerate}}
\newcommand{\een}{\end{enumerate}}

\def\frac#1#2{{{#1}\over {#2}}}
\def\gsim{\mathrel{\rlap{\lower4pt\hbox{\hskip1pt$\sim$}}
    \raise1pt\hbox{$>$}}}         
\def\lsim{\mathrel{\rlap{\lower4pt\hbox{\hskip1pt$\sim$}}
    \raise1pt\hbox{$<$}}}         

\newcommand{\dat}{\mathrm{dat}}

\newcommand{\draft}[1]{}

\def\lapprox{\lower .7ex\hbox{$\;\stackrel{\textstyle <}{\sim}\;$}}
\def\gapprox{\lower .7ex\hbox{$\;\stackrel{\textstyle >}{\sim}\;$}}

\numberwithin{equation}{section}
\numberwithin{figure}{section}
\numberwithin{table}{section}

\usepackage{tabularx}
\newcolumntype{C}[1]{>{\centering\arraybackslash}p{#1}}

\usepackage{amsmath}
\usepackage{amsfonts}
\usepackage{amssymb}
\usepackage{dsfont}
\usepackage{pifont}
\usepackage{booktabs}
\usepackage{graphicx}
\usepackage{epstopdf}
\usepackage{epsfig}
\usepackage{framed}
\usepackage{makeidx}
\usepackage{hyperref}
\usepackage{placeins}
\usepackage[font=small,labelfont=bf]{caption}

\newcommand{\nreps}{n_{\rm rep}}

\newcommand{\repind}{k}

\newcommand{\repchis}{{\chi^2}^{(\repind)}}
\newcommand{\cenchis}{{\chi^2}^{(\repind,\,c)}}
\newcommand{\cencenchis}{{\chi^2}^{(0,\,c)}}

\begin{document}
\newgeometry{top=1.5cm,bottom=1.5cm,left=1.5cm,right=1.5cm,bindingoffset=0mm}
\begin{figure}[h]
  \includegraphics[width=0.32\textwidth]{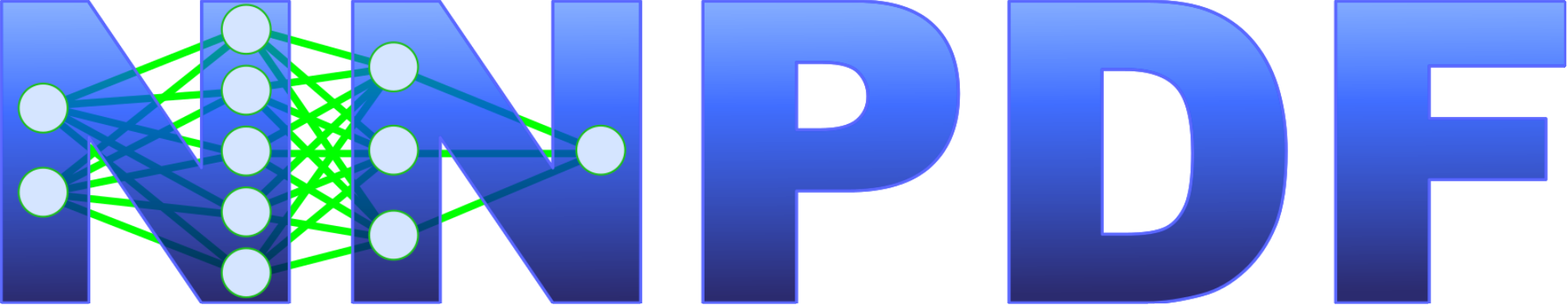}
\end{figure}
\vspace{-2.0cm}
\begin{flushright}
Edinburgh 2022/19 \\
TIF-UNIMI-2022-21\\
\end{flushright}
\vspace{0.3cm}

\begin{center}
  {\Large \bf 
  Response to ``Parton distributions need representative sampling''}
\vspace{1.1cm}

   {\small
     {\bf  The NNPDF Collaboration:}
\\[0.2cm]
 Richard D. Ball,$^{1}$
 Juan Cruz-Martinez,$^{2}$
 Luigi Del Debbio,$^{1}$
Stefano Forte,$^{3}$\\[0.1cm]
Zahari Kassabov,$^{4}$
Emanuele R. Nocera,$^{5}$
Juan Rojo,$^{6,7}$
Roy Stegeman,$^{1}$
Maria Ubiali$^{4}$\\[0.1cm]
 }

  \vspace{0.7cm}

  {\it \small

  ~$^1$The Higgs Centre for Theoretical Physics, University of Edinburgh,\\
    JCMB, KB, Mayfield Rd, Edinburgh EH9 3JZ, Scotland\\[0.1cm]
   ~$^2$ CERN, Theoretical Physics Department, CH-1211 Geneva 23, Switzerland\\[0.1cm]
     ~$^3$Tif Lab, Dipartimento di Fisica, Universit\`a di Milano and\\
    INFN, Sezione di Milano, Via Celoria 16, I-20133 Milano, Italy\\[0.1cm]
    ~$^4$DAMTP, University of Cambridge, Wilberforce Road,
    \\ Cambridge, CB3 0WA, United Kingdom\\[0.1cm]
 ~$^5$ Dipartimento di Fisica, Universit\`a degli Studi di Torino and\\
   INFN, Sezione di Torino, Via Pietro Giuria 1, I-10125 Torino, Italy\\[0.1cm]
    ~$^6$Department of Physics and Astronomy, Vrije Universiteit, NL-1081 HV Amsterdam\\[0.1cm]
 ~$^7$Nikhef Theory Group, Science Park 105, 1098 XG Amsterdam, The Netherlands\\[0.1cm]
 }

\vspace{1.0cm}

{\bf \large Abstract}

\end{center}

We respond to the criticism raised by Courtoy et al.~\cite{Courtoy:2022ocu}, in
  which the faithfulness of the NNPDF4.0 sampling is questioned
  and an under-estimate of the NNPDF4.0 PDF uncertainties is
  implied. We list, correct, and clarify in detail a number of 
  inaccurate or misleading claims that are made in
  Ref.~\cite{Courtoy:2022ocu}. Specifically, we explain and explicitly
  demonstrate why the central value of the PDF distribution does not generally coincide  with
  the absolute minimum of the $\chi^{2}$ to the data. We examine some
  PDFs that have been constructed in the above study and claimed to be ``good
  solutions'': we show that similar PDFs are found with the NNPDF methodology,
  but with very low probability.

\clearpage

\tableofcontents

\clearpage

\definecolor{amethyst}{rgb}{0.6, 0.4, 0.8}

\section{Introduction}
  \label{sec:Intro}

Progress in the understanding and testing of the Standard Model is increasingly
relying on precision measurements, which may reveal statistically significant
discrepancies between theory and experiment. The increase in luminosity at the
LHC, together with results from dedicated experiments, will drive a significant
increase in the precision of experimental results. It has been emphasized
multiple times that theory predictions need to match the forthcoming
experimental precision. A robust framework to estimate the statistical and
systematic errors in the determination of PDFs is one of the necessary
ingredients in this program. 

A recent publication~\cite{Courtoy:2022ocu} advocates a new technique for
sampling in the space of PDFs, or possibly for constructing PDFs.
Unfortunately, the statistical foundations of
this recipe are unclear and may lead to incorrect
conclusions. In what follows, we briefly summarize the challenges involved in
the extraction of PDFs, focusing on the statistical formulation, and point out
some of the problems with the claims in~\cite{Courtoy:2022ocu}.

The plan of this note is the following: In Sect.~2, after reviewing
the meaning of a PDF determination, we summarize how
PDFs  can be represented by Monte Carlo samples, the way in which the Monte
Carlo sample must be constructed, and show that NNPDF4.0 replicas
provide a representative sampling. A reader who is familiar with statistics 
and is mostly interested in the rebuttal can go directly to Sect.~3, where we 
list and correct a number of misleading or false statements that are made in
Ref.~\cite{Courtoy:2022ocu}. In Sect.~4 we examine some PDFs
constructed in~\cite{Courtoy:2022ocu}, we show that PDF replicas having
similar features can be obtained with the NNPDF4.0 methodology, but
that these PDFs correspond to overlearning, thus have a low probability.

In Appendix~\ref{sec:RegVal} we summarize the closure and future test indicators that have been used in
order to validate NNPDF4.0 PDFs, and a metric that can be used to
measure overlearning.

\section{Statement of the problem}
  \label{sec:StatementProb}

The determination of PDFs from a (necessarily finite) set of data is a
classical example of an inverse problem; the {\em functions} that we
are trying to reconstruct are elements of infinite-dimensional spaces,
and hence the problem is clearly under-determined and its solution 
will depend on the assumptions we make. 

\paragraph{Prior and posterior.} A Bayesian approach is well suited to tackle
these problems: the assumptions we make about the solution are encoded in a {\em
prior} probability measure in the space of functions. The probability measure is
then updated, yielding a so-called {\em posterior} distribution, using the
available data. Our knowledge about the solution of the inverse problem is 
encoded in the posterior~\cite{Giele:2001mr}.

In the particular case of PDFs, we are interested in functions
$f_k(x)$ where the index $k$ spans the set of PDFs in some chosen basis. 
For instance in the evolution basis used for NNPDF fits:
\begin{align}
  \label{eq:PDFList}
  f_k \in \left\{
    V,V_3,V_8,T_3,T_8,T_{15},\Sigma,g
  \right\}\, .
\end{align}
We refer to Ref.~\cite{NNPDF:2021njg} for a detailed description
of the flavor combinations that enter in the evolution basis.

The observables delivered by experiments can be expressed as convolutions
of these PDFs with coefficient functions, as dictated by factorization
theorems. A simple example of such a convolution is the factorization of the
DIS non-singlet structure function, see e.g. Ref.~\cite{DelDebbio:2007ee}.
In the case of a global fit, many observables corresponding to a large number of
different processes are combined in order to determine all the PDFs in the basis
in Eq.~(\ref{eq:PDFList}). Note that PDFs are defined, and can be extracted from
data, only within a well-defined theoretical framework (e.g. using perturbative
QCD in a given scheme and at a given order of perturbation theory). These
choices, together with any of the theoretical assumptions such as integrability,
positivity and sum rules, determine the prior for the probability measure.

Some parametrization for the functions $f_k$ needs to be chosen, so that the
problem can be cast in terms of a {\em finite} number of parameters. This is
broadly accepted by all collaborations that perform PDFs fits, albeit with 
different choices of the parametrization. Henceforth the finite-dimensional
space of parametrized functions is denoted by $\mathcal F$. In the NNPDF
approach, the parton distributions are parametrized using a single fully
connected deep neural network, so that the set of parameters corresponds to the
thresholds and weights of the network as explained in Ref.~\cite{NNPDF:2021njg}. The
set of parameters is traditionally denoted as
$\mathbf{\theta}$.

Bayes theorem allows us to write the posterior probability, given a dataset $D$:
\begin{align}
  \label{eq:BayesSol}
  p(\mathbf{\theta}|D,\mathcal H) \propto p(D|\mathbf{\theta},\mathcal H)
  p(\mathbf{\theta}|\mathcal H)\, ,
\end{align}
where $p(\mathbf{\theta}|\mathcal H)$ is the prior probability distribution,
i.e.\ the probability distribution in the space of parametrized functions {\em
before} being given any experimental results. The prior probability encodes all
the theoretical knowledge about the PDFs, as discussed above, and depends on the
hyperparameters $\mathcal H$ of the neural networks.

\paragraph{Monte Carlo representation of the posterior.}
It is customary to
approximate the posterior probability distribution by its mode
\begin{align}
  \label{eq:MAP}
  \mathbf{\theta}^* = \arg\max_{\mathbf{\theta}}
  p(\mathbf{\theta}|D,\mathcal H)\, ,
\end{align}
called the MAP (Maximum A Posteriori) estimator, complemented with the fluctuations 
of the MAP estimator induced by the fluctuations in the data. 
As explained since the very
early publications~\cite{Forte:2002fg}, the posterior distribution in the NNPDF
approach is constructed by a Monte Carlo sampling procedure. An ensemble of artificial data
$D^{(k)}$, called {\em replicas}, is generated, which reproduce the statistical
distribution of the experimental data. For each replica, the MAP estimator is
evaluated, which yields an ensemble of replicas of sets of parameters
$\mathbf{\theta}^{*(k)}$. It is important to note that in
Eq.~(\ref{eq:BayesSol}), the first factor is the likelihood,
\begin{align}
  \label{eq:RepLike}
  p(D^{(k)}|\mathbf{\theta},\mathcal H)\, ,
\end{align}
which is related as usual to the $\chi^2$ of the artificial data generated {\em
for that given replica} -- a quantity that we will consistently denote
$\repchis$. This should not be confused with the $\chi^2$ of that
replica to the central data, that we will consistently denote as
$\cenchis$.

Eqs.~(\ref{eq:BayesSol},\ref{eq:MAP}) show that the
quantity that is maximized when computing the MAP estimator is the posterior
probability, which also includes the prior distribution for the PDFs. Each
fitted replica is the mode of the posterior distribution for an instance of the
data that is generated with the probability distribution dictated by the
experimental uncertainties. In this way, the probability distribution of the data
induces the probability distribution of the parameters $\mathbf{\theta}$ and through the
Monte Carlo sampling procedure, it generates a sample of the posterior
distribution. In order to propagate the distribution correctly, each replica in
the ensemble must have the same statistical weight. Indeed, it
has been shown analytically that this procedure yields the correct
posterior distribution in the case of linear fits to Gaussian
data~\cite{DelDebbio:2021whr}, so that the expectation value of a generic
function $O(\mathbf{\theta})$ of the parameters $\mathbf{\theta}$ can be
computed as
\begin{align}
  \label{eq:ProbProp}
  \int d\mathbf{\theta}\, p(\mathbf{\theta}|F,\mathcal H)\, O(\mathbf{\theta})
  = \frac{1}{\nreps} \sum_{\repind=1}^{\nreps}
  O(\mathbf{\theta}^{(\repind)}), \quad \mathrm{for} \quad \nreps\rightarrow\infty.
\end{align}

The important points to keep in mind are:
\begin{enumerate}
  \item each PDF replica is obtained as the mode of the posterior distribution given
  one sample of the distribution of data;
  \item the posterior distribution is the product of the likelihood times the
  prior -- not just the likelihood;
  \item each replica has the same statistical weight;
  \item the $\chi^2$ of a replica to the central value of the experimental data
  is not a measure of the posterior probability of that replica, since the contribution of the prior is missing;
  \item outliers in the distribution of fitted replicas are typically good fits
  to some unlikely fluctuation of the data.
\end{enumerate}

\paragraph{Regularization.}
It is important to understand that even though each replica is obtained by
optimizing  $\repchis$, it does not correspond to the absolute minimum of this
quantity. Indeed, this absolute minimum generally corresponds to an overfitted
or overlearned solution. An overfitted solution is a solution that does not
generalize correctly the information contained in the specific instance of the
data. This means that, rather than fitting the true underlying law, it fits
particular features of the input data set or of the specific data instance
(such as a random statistical fluctuation). In the NNPDF methodology, such
overfitting is avoided through carefully constructed regularization methods.

First, before performing
the final fit, the methodology is hyperoptimized~\cite{Carrazza:2019mzf} through a K-folding algorithm~\cite{Forte:2020yip},
that  guarantees that choices related to the minimization algorithm are not
tuned to the specific datasets, thereby guaranteeing that the specific datasets
are not overfitted. Then, during the fitting itself, by using a cross-validation
algorithm, in which the minimization process is stopped before reaching the
absolute minimum using a training/validation split of the data. Specifically,
the minimization algorithm only receives as input the training part of the data,
while the validation part of the data is used as a control set. The final result
of a fit corresponds to the instance with lowest $\chi^2$ as defined with
respect to the validation part of the data. This ensure that any noise present
in the data is not inferred by the neural network. 
The training set (which in {\tt NNPDF4.0}~\cite{NNPDF:2021njg} is taken to be a
random set of 75\% of the points per dataset) is used to build the likelihood
function. The hyperoptimized  methodology then employs a gradient-descent-based
algorithm to take (for each replica) the path that maximizes the posterior
probability. Note that the settings of this algorithm are determined
through the hyperoptimization itself. Note also that the algorithm
partly depends on a random state that changes on a replica-by-replica
basis. The fit is stopped when the validation $\chi^{2}$ stops improving,
regardless of the value of the training $\chi^{2}$.

A number of tests of the robustness of the procedure have been performed
extensively by NNPDF: namely closure
tests~\cite{NNPDF:2014otw,DelDebbio:2021whr} that check sampling in the data
region and future tests~\cite{Cruz-Martinez:2021rgy} that check sampling in the
extrapolation region. These tests are briefly reviewed in
Appendix~\ref{sec:RegVal}.

We illustrate the outcome of this procedure when correctly implemented
by showing probability contours for the final replica distribution from the
 NNPDF4.0~\cite{Ball:2021dab} PDF determination.
Of course, a PDF set is a set of functions, so confidence levels for it
should be shown in a space of functions~\cite{Giele:2001mr}, which is
difficult to visualize.
We can instead consider a projection
of the PDF on a finite dimensional space. In order to make 
contact with the discussion in the following Sections, we choose 
a two-dimensional space of LHC cross sections, that of the  $Z$ and
Higgs total production 
cross section (ZH plane, henceforth), a choice that is also made in
Ref.~\cite{Courtoy:2022ocu }.
This is a useful  choice in that the Higgs
cross section is gluon driven and the $Z$ cross section is
quark-driven: so points in the ZH plane can be interpreted in terms of
the size of the quark and gluon luminosities.
Cross sections are computed as in~\cite{NNPDF:2021njg}.
In particular, partonic cross sections accurate to next-to-leading order (NLO)
in the strong coupling are convoluted with PDFs accurate to
next-to-next-to-leading order (NNLO). A center-of-mass
energy of 14 TeV is assumed, and cross sections are integrated in the fiducial
phase space specified in Sect.~9.2 of~\cite{NNPDF:2021njg}.
Contours in this plane provide a  test of the fact that the PDFs are correctly
sampled, given that the cross sections depend on several different combinations
of PDFs, evolved to the appropriate scale and convoluted over $x$ with hard
cross sections.

\begin{figure}[t]
  \center
  \includegraphics[width=0.6\textwidth]{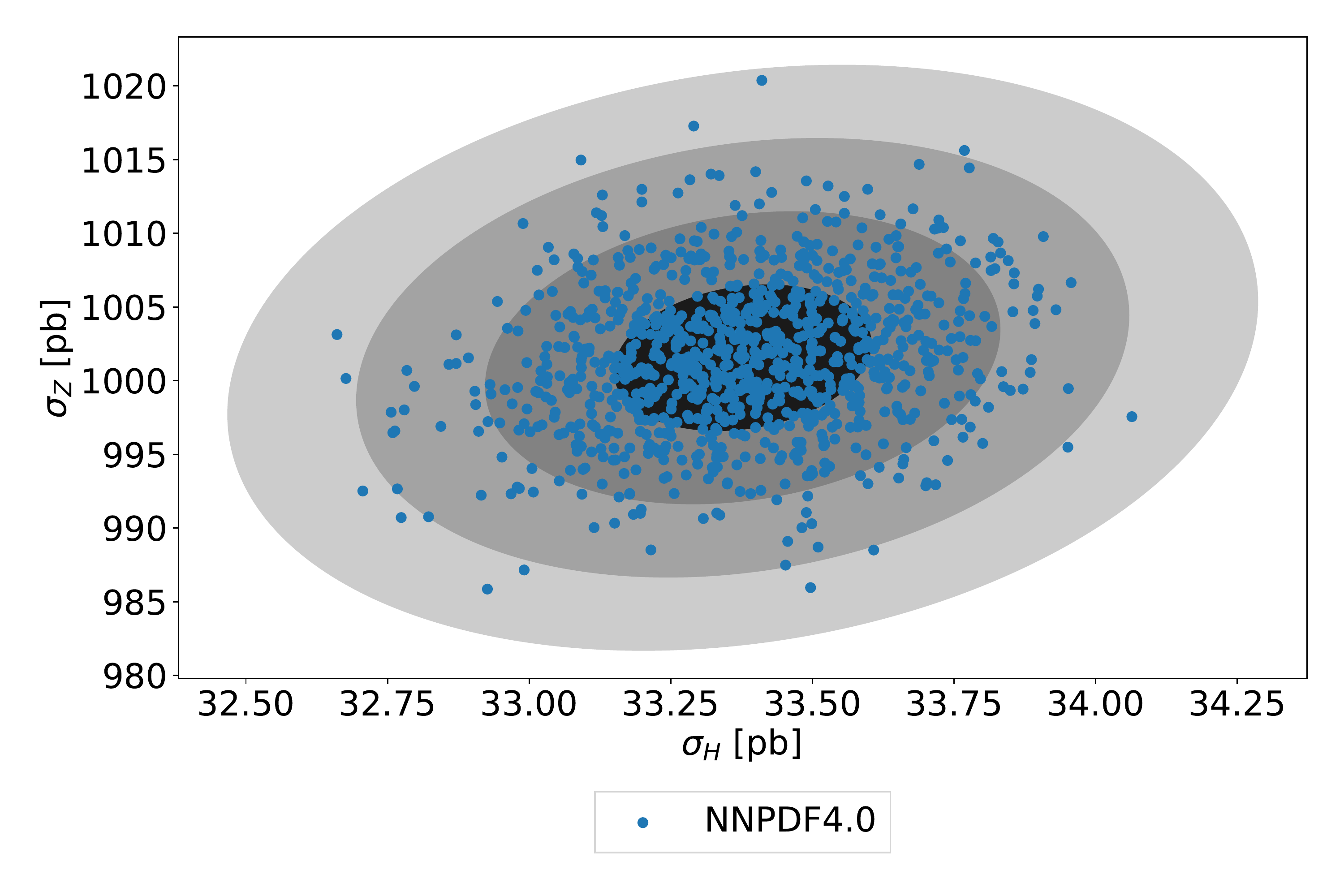}
  \caption{The NNPDF4.0 PDF replicas (1000 replica sample) in the
    $\sigma_Z$-$\sigma_H$ plane (referred to as ZH in the main
    text). The two, three  and four $\sigma$ contours are shown. 
}
  \label{fig:nnpdfmain}
\end{figure}

In Fig.~\ref{fig:nnpdfmain} we show the results for 1000
NNPDF4.0 PDF replicas in the ZH plane, along with two, three and four
$\sigma$ contours. 

The replicas are distributed as expected given that the underlying distribution
of the experimental uncertainties is Gaussian and it is expected (and
has explicitly shown~\cite{gaussian}) to lead to a Gaussian
distribution of physical predictions. Indeed, more replicas are concentrated at
the center and less in the tails, with no local distortions or clusters.  For
example, the two dimensional three $\sigma$ contour corresponds to a $98.9\%$
confidence level, so one would expect  11 NNPDF4.0 replicas outside the three
$\sigma$ contour, to be compared to  14 found in Fig.~\ref{fig:nnpdfmain} in
good agreement with the expected value. This result displays no evidence for
sampling bias in the NNPDF4.0 replica sample and instead confirms that the
replica sample is representative of the  probability distribution in the ZH
plane.

\FloatBarrier

\section{Inaccurate or misleading claims}
\label{sec:misleading}

In this section we discuss a number of misleading, inaccurate or false
claims that are made in Ref.~\cite{Courtoy:2022ocu}. These claims may
suggest that the PDF uncertainty on key LHC cross sections obtained with the
NNPDF4.0 PDF sets is underestimated.

First, we list individual claims and comment on them one by one. Then, we
discuss and correct a misconception on the meaning of the $\chi^2$
figure of merit that appears to underlie much of the discussion in Ref.~\cite{Courtoy:2022ocu}.

\begin{itemize}

\item In Section II it is stated that ``{\it analyses that fit a large number of
flexible functions using a modest number of fitted replicas might fail at
finding all possible solutions due to a sampling bias}''. It is then
  stated that  ``{\it dense sampling of a high-dimensional volume requires an exponentially growing number $N_p$ of replicas, such as
$2^{N_{\rm par}} \sim 10^{30}$  for $N_{\rm par}=100$}''.

This is misleading, as it suggests that a similar number of replicas is needed
for sets of PDFs. As discussed in Sect.~\ref{sec:StatementProb}, PDF fitting is
an example of an inverse problem: the aim is to find a {\it posterior}
probability of $f$ given the data {\bf D}. In the NNPDF approach, the MC
ensemble is not a random sampling, but rather an {\it importance sampling} of
the PDF space ${\cal F}$. As a result the number of replicas needed to obtain a
faithful representation in the PDF space does not require an exponentially
growing number of replicas. A replica sample with
$\nreps \sim 1000$ is sufficiently large to reproduce the correlations of the
experimental data to per-cent accuracy and to determine the 
the same accuracy. Note that this well-known fact is at
the basis of  the PDF4LHC15~\cite{Butterworth:2015oua} and
PDF4LHC21~\cite{PDF4LHCWorkingGroup:2022cjn} combinations, and its correctness
has been verified a posteriori several times in the construction of these sets,
e.g. by checking that samples of replicas of this size accurately represent
prior underlying sets of Hessian PDFs~\cite{Watt:2012tq,Hou:2016sho}. For
example, even with PDF sets with a relatively small number of parameters such as
CT18 or MSHT20, if the number  of replicas required in order to obtain a
faithful representation did scale as $2^{N_{\rm par}}$, then a number of
replicas of order $10^6-10^9$ would be required for a faithful representation,
instead of the $300$ used and shown to be adequate for the PDF4LHC combinations.
Conversely, most information contained in a replica sample can be represented
with a small number of Hessian parameters regardless of the size of the
sample~\cite{Carrazza:2016htc} suggesting a larger sample does not add
significantly more information.

\item In Section III it is stated that the NNPDF4.0
publication~\cite{NNPDF:2021njg} ``{\it interchangeably uses two forms of
$\chi^2$ as the figure-of-merit, called “$t_0$” and “exp”, which differ in
implementation of experimental systematic uncertainties.}''

This statement is false. The only figure of merit used by NNPDF for
minimization is the $\chi^2_{t_0}$.

This figure of merit,  used for minimization since the first NNPDF global
set~\cite{Ball:2009qv} is designed to avoid the so-called D'Agostini
bias~\cite{D'Agostini:642515} that would ensue in the presence of multiplicative
uncertainties (such as the luminosity uncertainty) if the covariance matrix as
published by experimental collaboration were used for minimization. The
experimental covariance matrix is written as
\begin{equation}
  \label{eq:cov}
  \left(\operatorname{cov}_{\exp }\right)_{i j}=\delta_{i j}\sigma_i^{\text {(uncorr) }}\sigma_j^{\text {(uncorr) }}+\left(\sum_{m=1}^{N_{\rm mult}} \sigma_{i, m}^{(\rm norm)} \sigma_{j, m}^{(\rm norm)}+\sum_{l=1}^{N_{\rm corr}} \sigma_{i, l}^{(\rm corr)} \sigma_{j, l}^{(\rm corr)}\right) D_i D_j,
\end{equation}
where the indices $i,j$ label the datapoints, $\sigma_i^{\text {(uncorr) }}$ are the uncorrelated uncertainties obtained by adding the uncorrelated systematic uncertainties and statistical uncertainties in quadrature, $m$ runs over the $N_{\rm norm}$ multiplicative normalization uncertainties, $\sigma_{i,m}^{\text {(norm) }}$, and $l$ runs over the $N_{\rm corr}$ other correlated systematic uncertainties, $\sigma_{i,l}^{\text {(corr) }}$. Finally, $D_i$ are the measured central values.

The $t_0$ covariance matrix is then defined as
\begin{equation}
  \label{eq:covt0}
  \left(\operatorname{cov}_{t_0 }\right)_{i j}=\delta_{i j}\sigma_i^{\text {(uncorr) }}\sigma_j^{\text {(uncorr) }}+\sum_{m=1}^{N_{\rm norm}} \sigma_{i, m}^{(\rm norm)} \sigma_{j, m}^{(\rm norm)}T_i^{(0)} T_j^{(0)}+\sum_{l=1}^{N_{\rm corr}} \sigma_{i, l}^{(\rm corr)} \sigma_{j, l}^{(\rm corr)}D_i D_j,
\end{equation}
where $T^{(0)}_i$  is a theoretical prediction for the $i$-th data
point evaluated using a $t_0$ input PDF.

In the  $t_0$ method~\cite{Ball:2009qv}, PDFs are determined by optimizing
\begin{equation}
  \label{eq:chi2t0}
\chi^2_{t_0}=\sum_{i, j}^{N_{\mathrm{dat}}}\left(T_i-D_i\right)\left(\operatorname{cov}_{t_0}^{-1}\right)_{i j}\left(T_j-D_j\right).
\end{equation}
This leads to a set of PDFs that is then used to iteratively compute the
theoretical predictions $T_i^{(0)}$ that are needed for the construction of a
new $t_0$ covariance matrix, which is then used for a new PDF determination.
This procedure is iterated until convergence.

Note that all PDF determinations use similar procedures for the
treatment of multiplicative uncertainties, since, as mentioned, if the
experimental
$\chi^2$ were used for minimization, biased results would be
obtained. The $\chi^2$ definitions used by different groups in order
to avoid the D'Agostini bias were
 benchmarked in
Ref.~\cite{Ball:2012wy},  where in particular it was shown that the
$t_0$ definition is entirely equivalent to that used by the CT and
MSTW groups. 

The experimental $\chi^2$, which is obtained using the experimental
covariance matrix $\operatorname{cov}_{\exp}$ of Eq.~(\ref{eq:cov}) instead of
$\operatorname{cov}_{t_0}$ of Eq.~(\ref{eq:covt0}) in the definition of
Eq.~(\ref{eq:chi2t0}) is never used by NNPDF for fitting. It is only quoted for
the sake of comparison to $\chi^2$ values computed and quoted by experimental
collaborations.

Hence, any value of the experimental $\chi^2$, such as those
displayed in Figs.~2, 5 and 6 of Ref.~\cite{Courtoy:2022ocu}, is
meaningless if the purpose is to discuss fit quality, given that this
is not the figure of merit that is used for PDF
determination. Specifically, it is meaningless if the goal is
to show that there exists PDF combinations that display a better
$\chi^2$ than the one obtained in the  NNPDF4.0 fit.
In the sequel, when referring to $\chi^2$ values we will always mean
the $t_0$ $\chi^2$.

\item In Section III it is stated that ``{\it Each replica fit achieves a good
$\chi^2$ with respect to its fluctuated data set, while practically all MC
replicas have a very high $\chi^2$ (by hundreds of units) with respect to the
published (unfluctuated) data values. The individual MC PDFs are thus poor fits
to the published (unfluctuated) data set – but their average (called the
”central replica”, or ”replica 0”) agrees with the unfluctuated data much
better.}''

This statement is misleading, in that it suggests that the quality of the fit of
PDF replicas to the central unfluctuated data is a relevant measure of their fit
quality. This is incorrect. Each PDF replica is determined through a fit to its
data replica and therefore the fluctuations of the fits truly reflect the
fluctuations in the PDFs induced by the fluctuations in the data. 

\item In Section III.D it is stated that ``{\it in the current NNPDF
procedure, small data sets may acquire a larger effective $\chi^2$ weight,
because during the cross validation the small data sets may be included in their
entirety in the fitted sample and not divided between the fitted and control
samples. The resulting PDFs are then biased toward the smaller data sets.}''

The statement is false. As clearly stated in Ref.~\cite{NNPDF:2021njg} (see
Section 3.2.4) for all datasets (regardless of whether large or small) 75\% of
data enters the training set.  In NNPDF4.0, the training fraction was increased
from the value 50\%  used in previous NNPDF determinations based on the
observation that the dataset analyzed in the NNPDF4.0 fit is so wide that,
even with just 25\% validation, overlearning does not occur. This observation
was backed up by the closure test studies performed in
Ref.~\cite{DelDebbio:2021whr} and an explicit check using the $R_O$ metric of Eq.~(\ref{eq:overfitting_metric}).
\end{itemize}

\begin{figure}[htp]
    \center
    \includegraphics[width=0.8\textwidth]{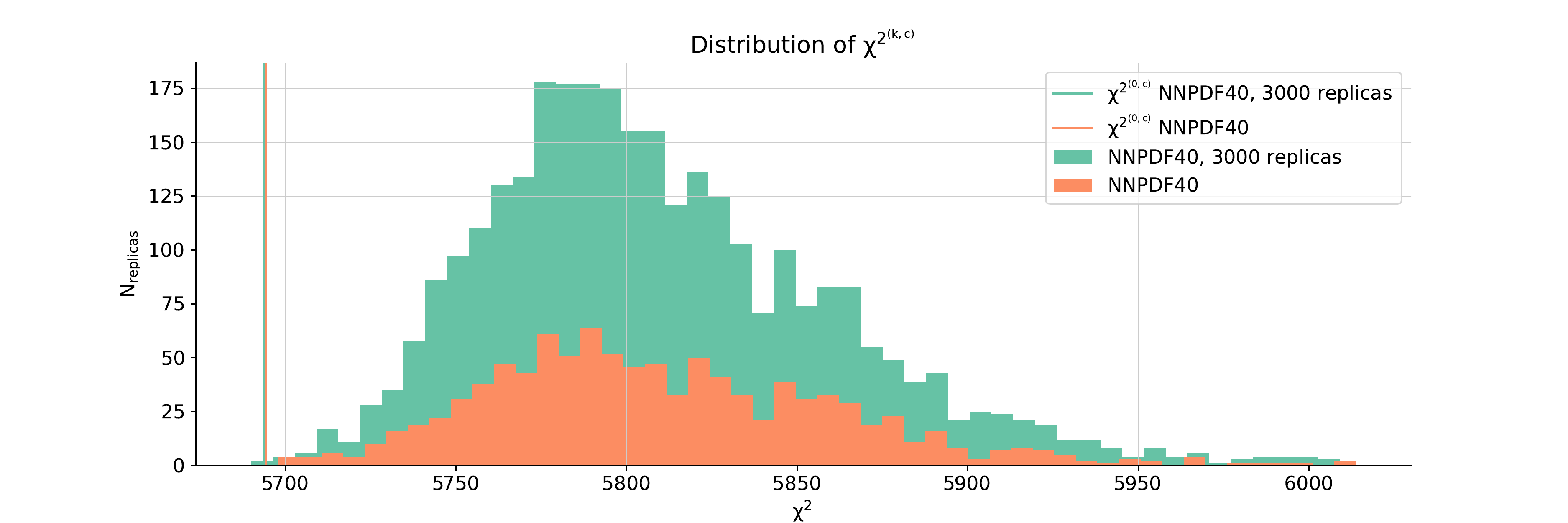}
    \caption{Histogram of the $\cenchis$ values (computed with the
      $t_0$ definition) for sets of 1000
      (red) or 3000 (blue) NNPDF
      replicas. The value $\cencenchis$ of the  central value,
      obtained as an average over the given replicas, is also shown in
      each case.}
    \label{fig:minchi1000}
\end{figure}

Finally, a clear misconception  emerges in~\cite{Courtoy:2022ocu}, which leads
to various confusing or wrong statements. Recall that in the NNPDF approach, the
central PDF is obtained as the expected value, i.e. as the average over the
replica sample. It is sometimes also referred to as replica 0 (because this is
how it is delivered through the LHAPDF6 interface~\cite{Buckley:2014ana}). We
denote the $\chi^2$ of this mean, replica 0  PDF to the central data as
$\cencenchis$. The misconception is that  $\cencenchis$ should be the absolute
minimum of  $\chi^{2}$. This is incorrect for several reasons that we recall
here:
\begin{itemize}
  \item As already mentioned, in order to avoid overfitting each PDF replica,
    $f^{(k)}$, to the corresponding data replica, $D^{(k)}$, an early stopping
    procedure is implemented. Hence a PDF replica does not correspond to the
    absolute minimum of $\repchis$. Likewise, it is easy to construct PDFs that
    provide an overfit to central data, such as the PDF shown in Fig.~3.7 of
    Ref.~\cite{NNPDF:2021njg}. These correspond to a lower value of $\chi^{2}$
    than the average of the replica sample.
  \item Whatever the value of the average obtained from a given sample, there
    always exists any number of PDFs with lower values, trivially because the
    average of any given replica sample is just an instance of a probability
    distribution. There will be an instance of the distribution that corresponds
    to the minimum of  the $\chi^{2}$ for the given distribution, but the
    probability of sampling exactly this instance is of course zero. These PDFs
    will be indistinguishable from the ones that are provided within PDF
    uncertainties.
  \item As already explained, NNPDF replicas are constructed as fits to data
    replicas, i.e. by minimizing $\repchis$, which is clearly not the same as
    $\cenchis$, where the latter measures the agreement to central data. This
    means that $\cenchis$ should have a
    probability distribution, that depends on the likelihood of individual data
    replicas. Hence, in particular, for a large enough replica sample there will
    always exist increasingly unlikely data replicas, leading to correspondingly
    unlikely PDF replicas, with arbitrarily high $\cenchis$. Conversely, some of
    them will be in the opposite tail of the distribution, such that their $\cenchis$ is
    lower than $\cencenchis$. This is illustrated in Fig.~\ref{fig:minchi1000},
    where we show the values of $\cenchis$ of each replica $k$ plotted in a
    frequency
    histograms with the $\chi^2$ of the central PDF marked as a vertical line.
    Results are shown for a sample of 1000 (red) and 3000 (blue) replicas. It is
    clear that for a large enough number of replicas there are some that have  a
    value of $\chi^{2}$ smaller than that of the average.
\end{itemize}

\section{The hopscotch PDFs}
\label{sec:hopscotch}

The main result presented in  Ref.~\cite{Courtoy:2022ocu} is the
construction of a set of PDFs by the so-called hopscotch (HS)
method. We first summarize the construction of these PDFs, then their possible
interpretation, and finally state some questions that they may
raise. We then present the answer to these questions.

The HS PDFs are constructed as follows: first, the Hessian version of the
NNPDF4.0 is used. This Hessian version has been constructed~\cite{NNPDF:2021njg}
using the {\tt mc2hessian}~\cite{Carrazza:2015aoa} code, which determines a
multigaussian projection of the probability distribution obtained from a given
replica sample with a desired number of Hessian eigenvectors by sampling the
replica probability distribution, performing a singular value decomposition of
the result, and retaining the eigenvectors with largest eigenvalues. The set
delivered in Ref.~\cite{NNPDF:2021njg} contains 50 eigenvectors, which is
sufficient to guarantee percent-level accuracy on the PDFs. Then, new PDFs are
sought for by moving along each Hessian eigenvector direction, monitoring the
$\chi^2$ value, and specifically looking for the lowest $\chi^2$ configurations.
These HS PDFs thus correspond to first, taking PDFs that correspond to
one-$\sigma$ deviations from the center of the given NNPDF4.0 replica sample,
and then performing linear combinations of them. Hence, they can be thought of
as linear combinations of NNPDF4.0 replicas --- though strictly speaking the
PDFs of which they are linear combinations are not NNPDF4.0 replicas, but rather
Hessian PDFs constructed out of the replica distribution. Note that these HS
PDFs are not a part of an ensemble of replicas and must therefore
considered as isolated PDF instances.
Indeed, as correctly stated in Ref.~\cite{Hou:2016sho}, replicas are
only useful as an ensemble.

Several of these HS PDFs are constructed by looking at and minimizing
the experimental $\chi^2_{\rm exp}$. As discussed in
Sect~\ref{sec:misleading}, $\chi^2_{\rm exp}$ is never used as a
figure of merit for PDF minimization, by either NNPDF or other fitting
group, since its minimization would lead to PDFs affected by
d'Agostini bias. Hence the HS PDFs constructed by looking at this
figure of merit are devoid of interest or significance and we will
ignore them: they only confuse the relevant issues.

\begin{figure}[t]
    \center
    \includegraphics[width=0.6\textwidth]{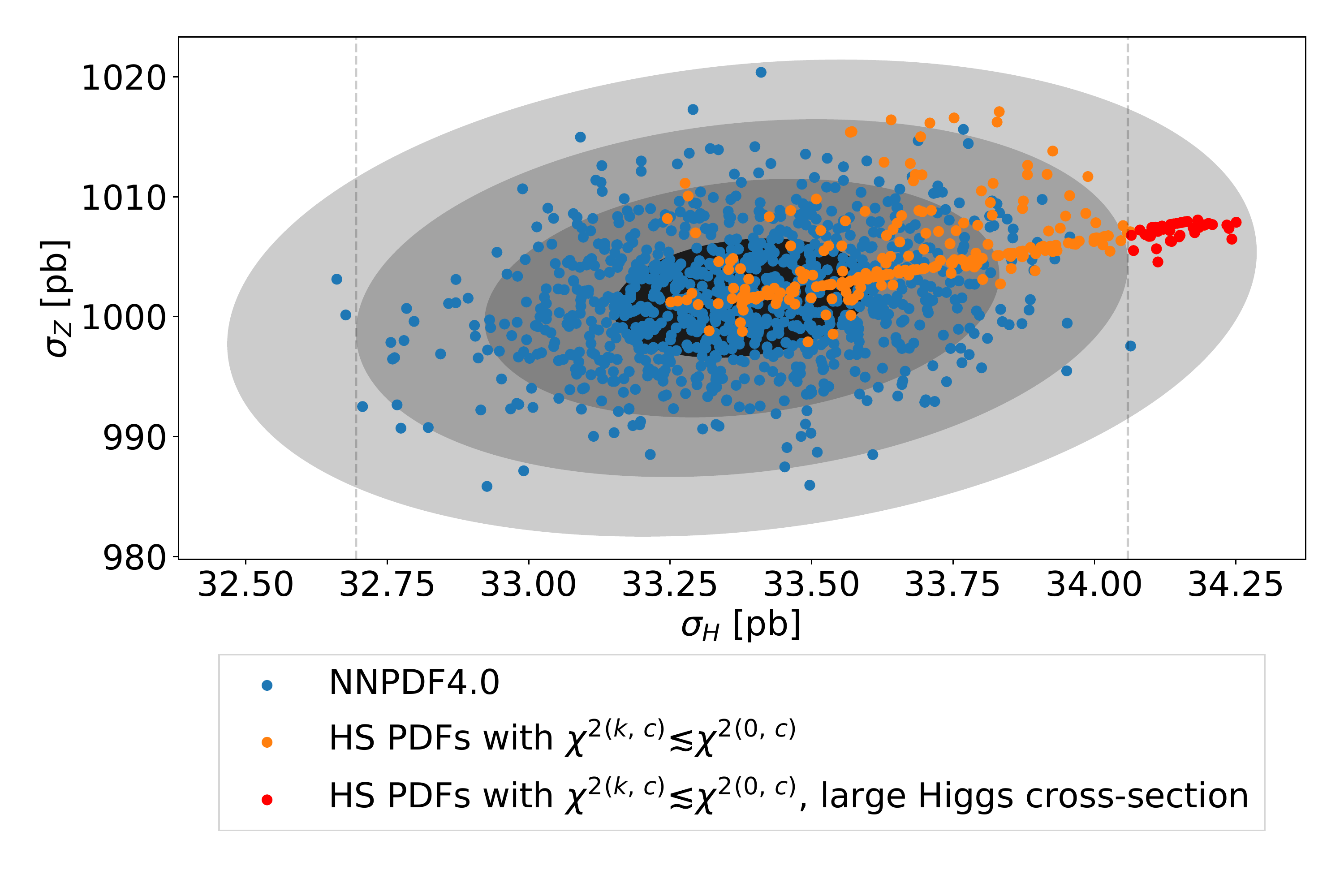}
    \caption{Same ad Fig.~\ref{fig:nnpdfmain}, but now also including
      a set of 300 HS PDFs with $\chi^2$ values
      similar to that of the NNPDF4.0 central value, 
      $\cenchis\lsim \cencenchis$ (orange and red points). The
      three $\sigma$ interval for  $\sigma_H$ is denoted by two vertical
      bands, and the HS PDFs that falls outside it are shown in red.
}
    \label{fig:hopscotchmain}
\end{figure}

Further HS PDFs are instead constructed looking at and minimizing the $t_0$
$\chi^2$ (that we are consistently denoting as $\chi^2$ for short)
and we discuss them now. Specifically, these PDFs are constructed by
minimizing $\cenchis$, namely the $\chi^2$ to central data --- unlike
NNPDF4.0 replicas that are determined minimizing $\repchis$, the
$\chi^2$ to a data replica.  Note that here  the index $k$ merely numbers the HS
PDFs, which as already stated are not an ensemble of replicas, but
rather isolated PDF instances.
Fig.~\ref{fig:hopscotchmain} we show the 300 HS PDFs that have values
of $\cenchis$ comparable or lower than $\cencenchis$, the value of the
central NNPDF4.0 PDF:
$\cenchis\lesssim\cencenchis$.  They are shown in the ZH plane,
superposed to the set of 1000 NNPDF4.0 replicas of Fig.~\ref{fig:nnpdfmain}.

It is manifest that whereas the NNPDF4.0 replicas provide a representative
sampling of the underlying probability distribution, the HS PDFs instead are
mostly cluster along one line in the ZH plane. These HS PDFs generally have
large values of the Higgs cross section, and a large number of them, denoted by
red points in Fig.~\ref{fig:hopscotchmain}, fall outside the three $\sigma$
interval for this cross-section as measured by the NNPDF4.0 replicas, denoted by
vertical lines in the figure. Because three $\sigma$ corresponds to $99.7\%$ in
one dimension, three  replicas are expected outside the three $\sigma$ interval
for the Higgs cross section, and  indeed three NNPDF4.0 replicas fall outside
it, two to the left and one to the right.

The smallest value of $\cenchis$ for the HS PDFs is about
$\cenchis-\cencenchis=-0.01$ (about 40 units if not normalized to the
number of datapoints as in Eq.~\eqref{eq:chi2t0}, with about 4000
datapoints). The standard deviation of $\chi^2$ is
$\sigma_{\chi^2}=\sqrt{2/N_{\rm dat}}\approx 0.02$, so these HS PDFs
have a  $\chi^2$ value 
that differs by half $\sigma$ from $\cencenchis$. It is
important here not to confuse this hypothesis-testing fluctuation of
the $\chi^2$ with the parameter-fitting criterion $\delta \chi^2=
1/N_{\dat }$~\cite{Collins:2001es}, which would instead lead one to
think that these PDFs are many standard deviations away. Indeed, 
$\cenchis$ values do not correspond to fitting a single fixed functional
form to the central data. Rather, as explained in
Section~\ref{sec:StatementProb} and repeatedly stated,
they correspond to fitting PDF replicas to data replicas. The
extremely flexible NNPDF parametrization (with about 800 free
parameters) explores the space of functional forms ----- indeed, it
has been shown that NNPDF fits to central data include a functional
component of the uncertainty ~\cite{NNPDF:2014otw,Ball:2021dab}, so
different PDF replicas should be viewed as instances of both different
underlying data and underlying functional forms, that are sampled
e.g. as the random state of the minimzation is varied in the replica
sample as explained in Section~\ref{sec:StatementProb}. Note finally that the
accuracy of the Hessian approximation to the probability distributions
of PDFs
on which the whole construction is based is at the percent level.
For all these
reasons, the HS PDFs should be viewed as PDFs that have about the same
value of $\cenchis$ as the NNPDF4.0 central value: i.e. such that
$\cenchis\approx\cencenchis$.  Because
the HS PDFs do not define a measure of probability, it is a priori unclear to
what extent the larger values of the Higgs cross section are incompatible with
the NNPDF determination and whether more NNPDF replicas would appear superposed
to them in the ZH simply increasing the size of the NNPDF replica sample. 
To clarify this point we study the following  two  questions:
\begin{enumerate}
\item Does the NNPDF4.0 methodology have difficulties in producing
  PDF replicas that look
  like the HS PDFs, and specifically that have large values of the
  Higgs cross-section, specifically outside the three $\sigma$
  interval and with $\sigma_H$ 
    bigger or much bigger than the mean?
    Or is there perhaps some reason (such as, for instance, an
    excessive stiffness  for some reason built in the NNPDF4.0
    methodology) that prevents such PDFs being obtained, so that e.g. there
    is a ``hard 
    wall''  along the $\sigma_H$ axis,  so that the NNPDF4.0 methodology
    cannot produce replicas with sufficiently large Higgs cross-section?
  \item If instead
     the NNPDF4.0 methodology can produce replicas with large
     $\sigma_H$, and/or replicas that look like the
     HS PDFs, why are these replicas so unlikely,
     despite having $\cenchis\sim\cencenchis$? As
      discussed in Sect.~\ref{sec:StatementProb}, there is nothing
      wrong with a PDF being unlikely despite having a low
      $\cenchis$, because the posterior distribution is the product of
      the likelihood times the prior, so a PDF could have high
      likelihood yet low prior probability. However, this then begs
      the question, why is the prior probability for these replicas so
      low?
\end{enumerate}

We now address these two questions in turn, in two dedicated
subsections.
Specifically we show that
\begin{enumerate}
  \item the NNPDF4.0 fitting methodology has no difficulty in
    producing replicas that look like the HS PDFs and/or PDFs
    leading to
    the largest values of $\sigma_H$ of the HS PDFs;
    \item the HS PDFs have low likelihood because they correspond to
      overfitted solutions.
\end{enumerate}

      \subsection{NNPDF4.0 replicas reproducing the HS PDFs}

      Fig.~\ref{fig:hopscotchmain} shows that NNPDF4.0 predicts a multigaussian
      distribution in the ZH plane.  Therefore, one expects that in order to get an
      arbitrarily large number of PDF replicas in any given region (specifically
      with large $\sigma_H$) it is  sufficient to
      increase the number of replicas by a sufficiently large amount. Yet it
      could be that this is not the case and that something makes it
      more difficult for the
      NNPDF4.0 methodology  to produce PDFs with large values of the
      Higgs cross-section. We can test whether this is the case in various ways.

      As a first test, we recall that NNPDF4.0 replicas are fitted to
      data replicas by minimizing $\repchis$. If for the NNPDF4.0
      methodology it is more difficult to fit replicas in the rightmost
      region of the ZH plane, then we should see a correlation between
      the value of $\repchis$ and the position of the replica in the
      ZH plane. Specifically, we should see that replicas with high
      Higgs cross-section also have a higher value of  $\repchis$
      
\begin{figure}[htp]
    \center
    \includegraphics[width=0.49\textwidth]{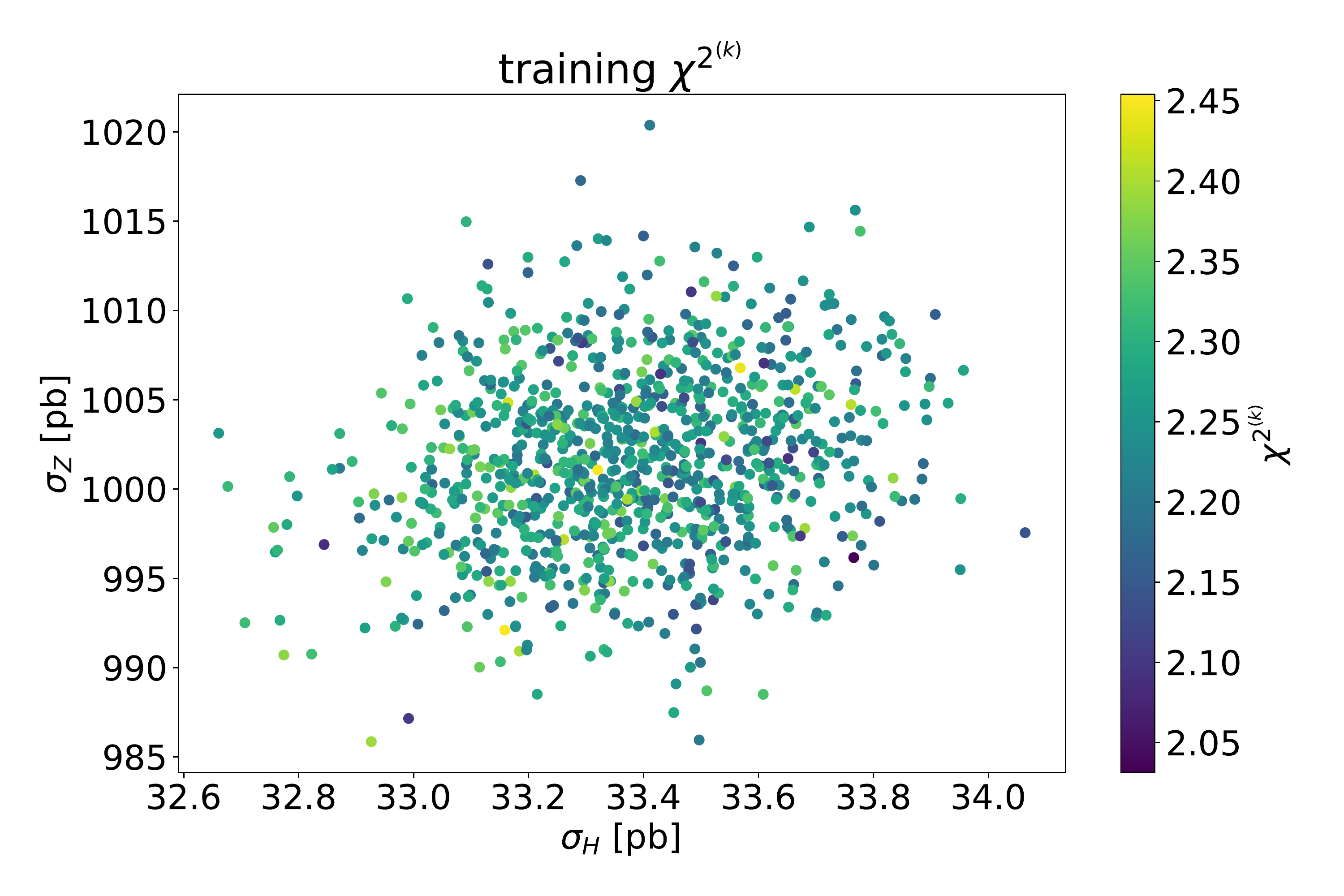}
    \includegraphics[width=0.49\textwidth]{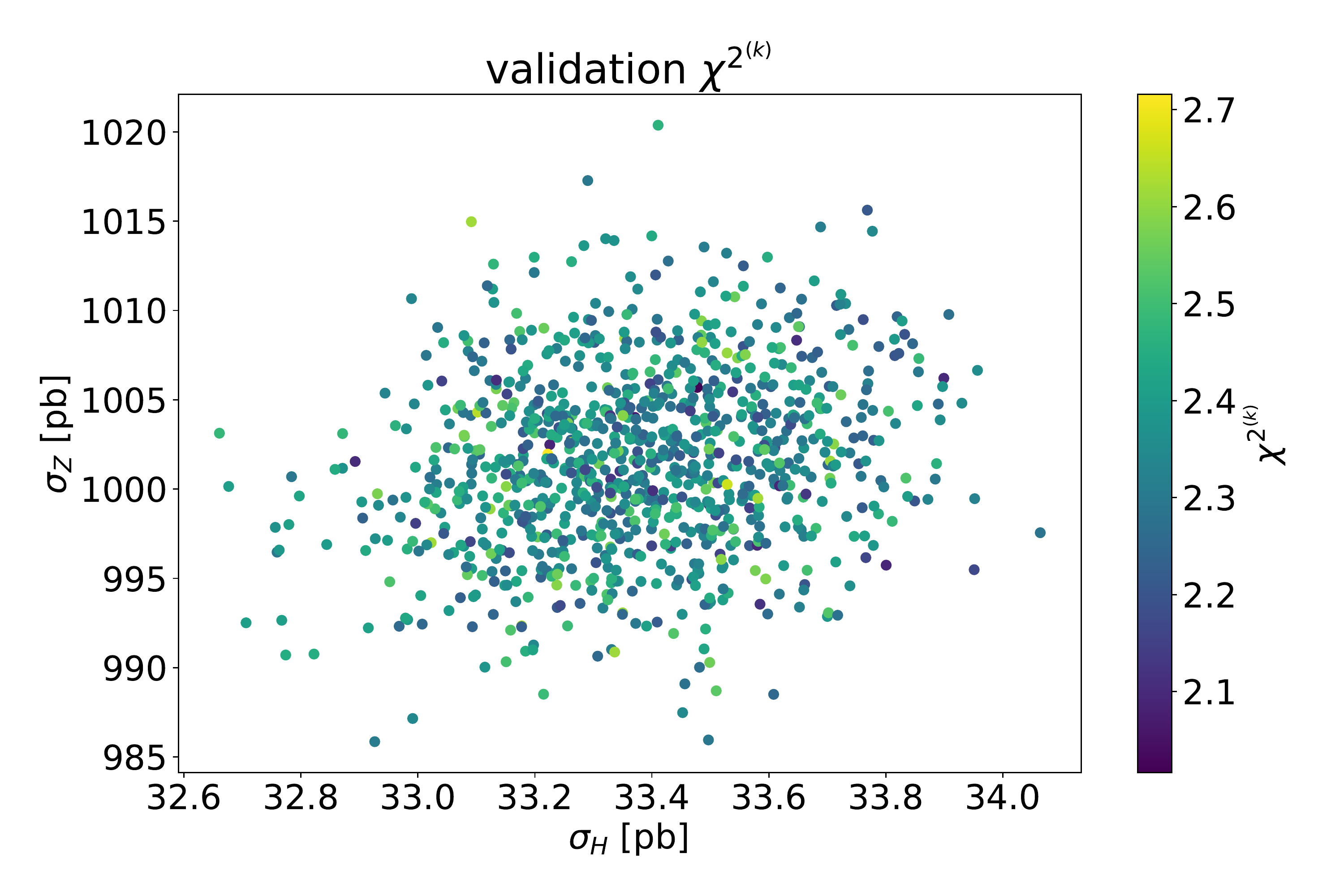}
    \caption{Scatter plot of the training (left) and validation
      (right) values of $\repchis$ for NNPDF4.0 replicas in the
      $\sigma_h-\sigma_Z$ plane. The value of $\repchis$ is shown as a color
      code.}
    \label{fig:chi2cloudtrvl}
\end{figure}

The values of $\repchis$ for both the training and validation data for
the NNPDF4.0 replicas are shown in Fig.~\ref{fig:chi2cloudtrvl}. It is
clear that there is no visible correlation between the position in the
ZH plane, specifically the position along the $\sigma_H$ axis,
and the value of the $\repchis$. In fact, the fit quality of
each PDF replica to its data replica is similar, and essentially
independent of the position in the ZH plane. This means that outlier
replicas are fitted equally well as replicas close to the center of
the distribution. Outlier replicas simply correspond to unlikely data
fluctuations. The NNPDF4.0 methodology has no difficulties in fitting
PDFs that correspond to
large (or small) values of the Higgs (or Z) cross-section.

\begin{figure}
  \center
  \includegraphics[width=0.6\textwidth]{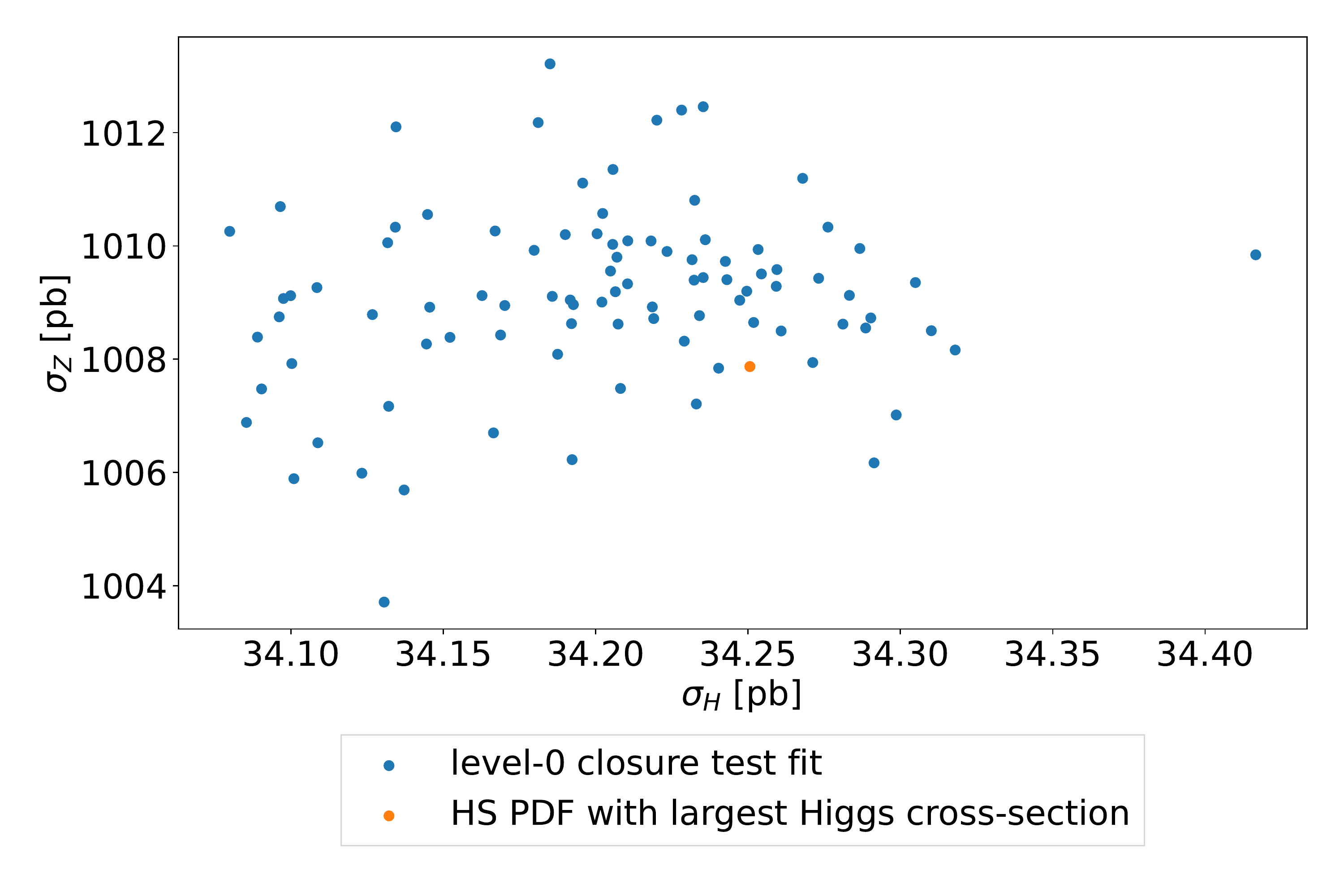}
  \caption{Scatter plot in the $\sigma_h-\sigma_Z$ plane of
    PDF replicas determined in level-0 closure test with the
     HS PDF with largest Higgs cross-section taken as underlying
     truth, shown as an orange point.}
  \label{fig:level0_scatterplot}
\end{figure}

As a second test, we check explicitly that we can fit the HS PDFs if we assume
them to be the underlying truth. To this purpose, we have performed a level-0
closure test~\cite{NNPDF:2014otw}. The closure test procedure is summarized in
Appendix~\ref{sec:RegVal}. Level-0 means that we have generated data with
zero uncertainty, assuming a given underlying true PDF, and then fitted these
data. Because the data are fitted at zero uncertainty, the fit can obtain
vanishing $\chi^2$. We have picked as an underlying true PDF the HS PDF that
gives the largest Higgs cross-section.  We have then fitted 100 PDF replicas to
it, with standard methodology (including training-validation split). We find
$\langle \repchis \rangle_{\rm tr}=0.03\pm0.01$, $\langle \repchis \rangle_{\rm
val}=0.04\pm0.02$, so indeed we reach a near-prefect fit. A scatter plot of
results in the ZH plane is shown in Fig.~\ref{fig:level0_scatterplot}. It is
clear that even though all the fits are equally good and fit the data perfectly
(with zero uncertainty) there is still a distribution of results, due to the
fact that of course the data do not determine the PDF completely. Each replica
can be thought of as a different equally good interpolation of the given data,
distributed in the ZH plane. Several of these results have values of $\sigma_H$
that are in fact larger than the input underlying truth. So there is surely no
``hard wall'', and  we must conclude that the NNPDF4.0 methodology has no
difficulty in producing replicas with large Higgs cross section and/or of
fitting the HS PDF exactly.
\begin{figure}
  \center
  \includegraphics[width=0.6\textwidth]{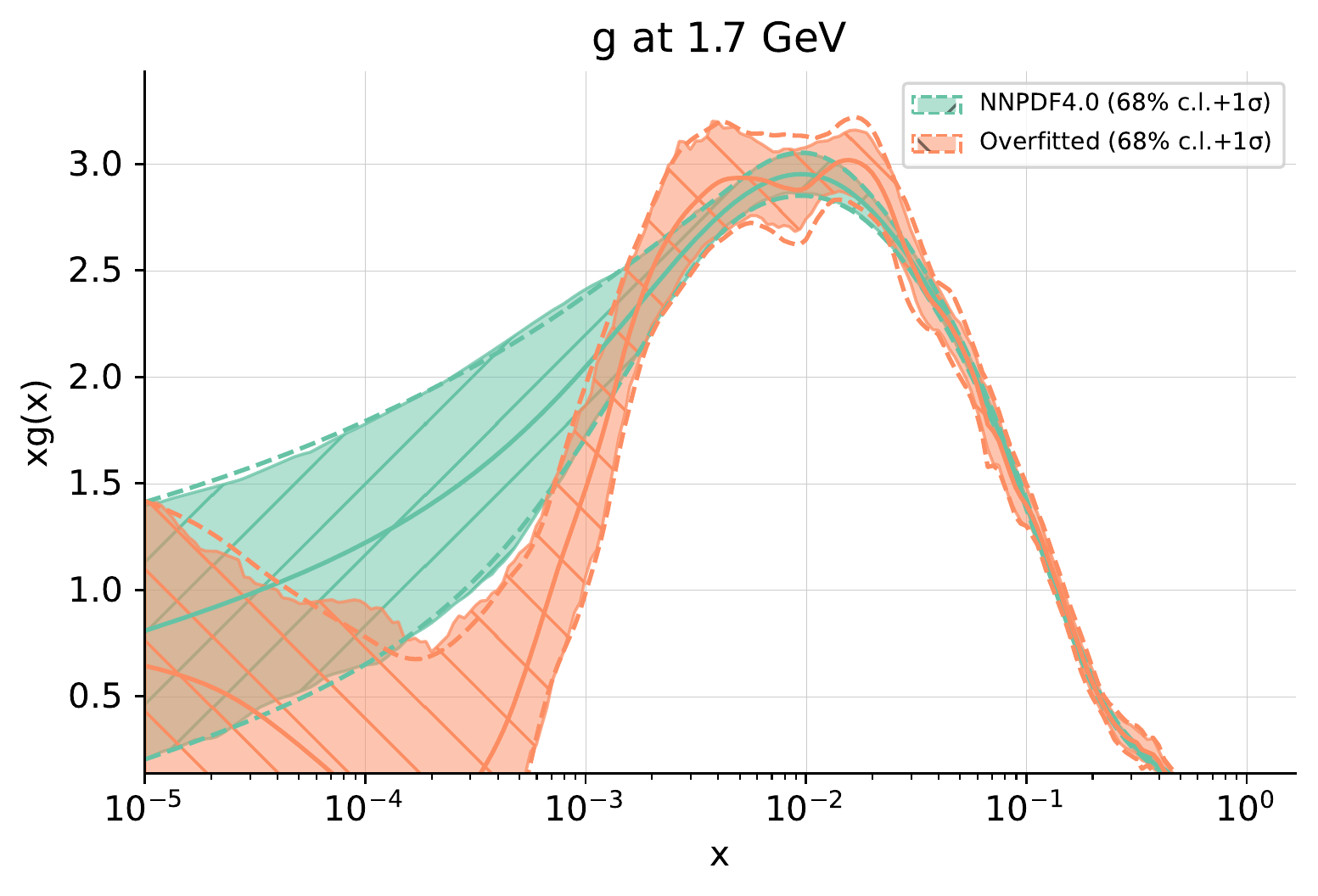}
   \caption{The gluon obtained in a overfitted PDF determination, in
     which the final $\cencenchis$ value is by about 0.8 smaller than
     that of the default NNPDF4.0.}
  \label{fig:overfit}
\end{figure}

\subsection{The HS PDFs and overfitting}

\begin{figure}
  \center
  \includegraphics[width=0.45\textwidth]{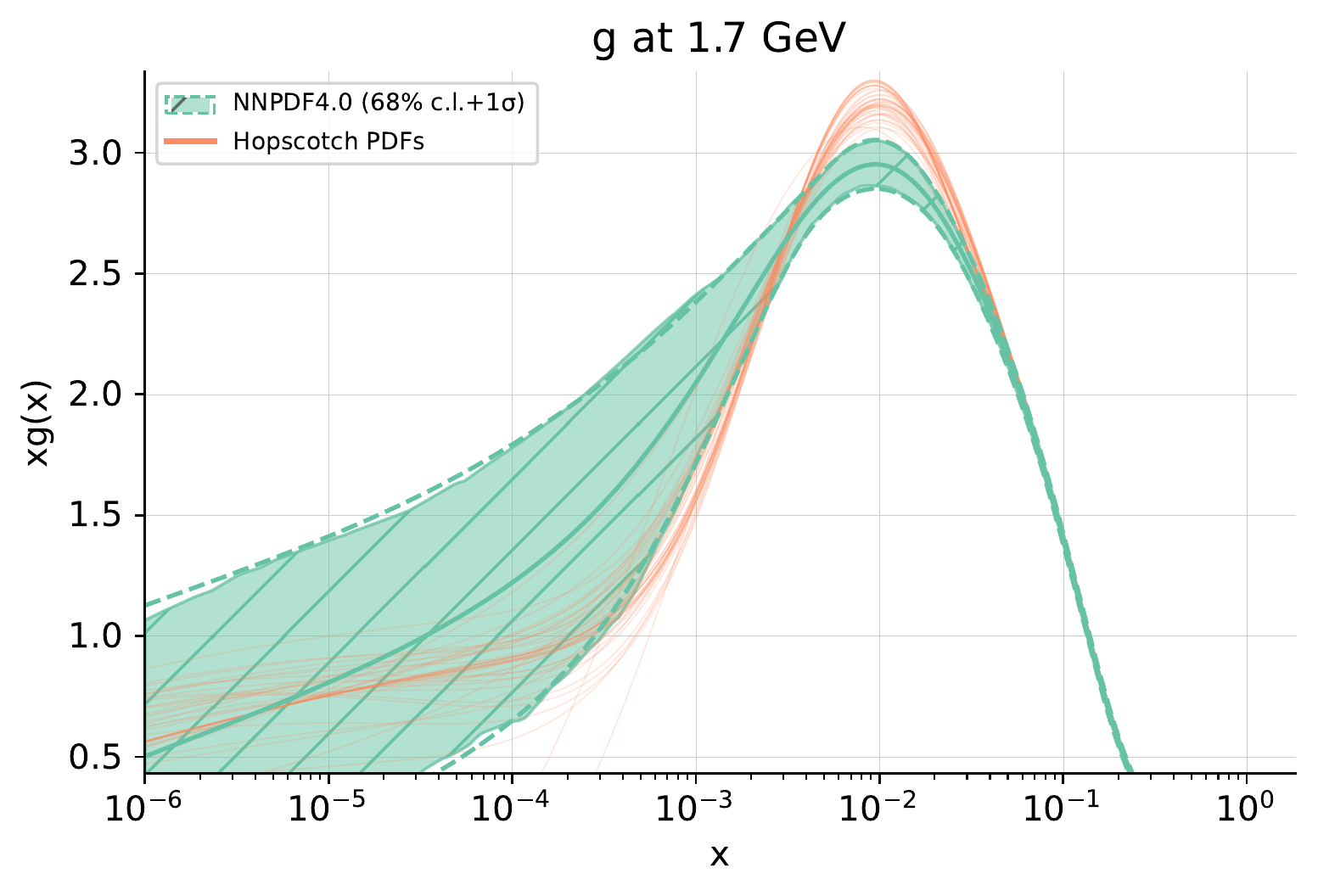}
  \includegraphics[width=0.45\textwidth]{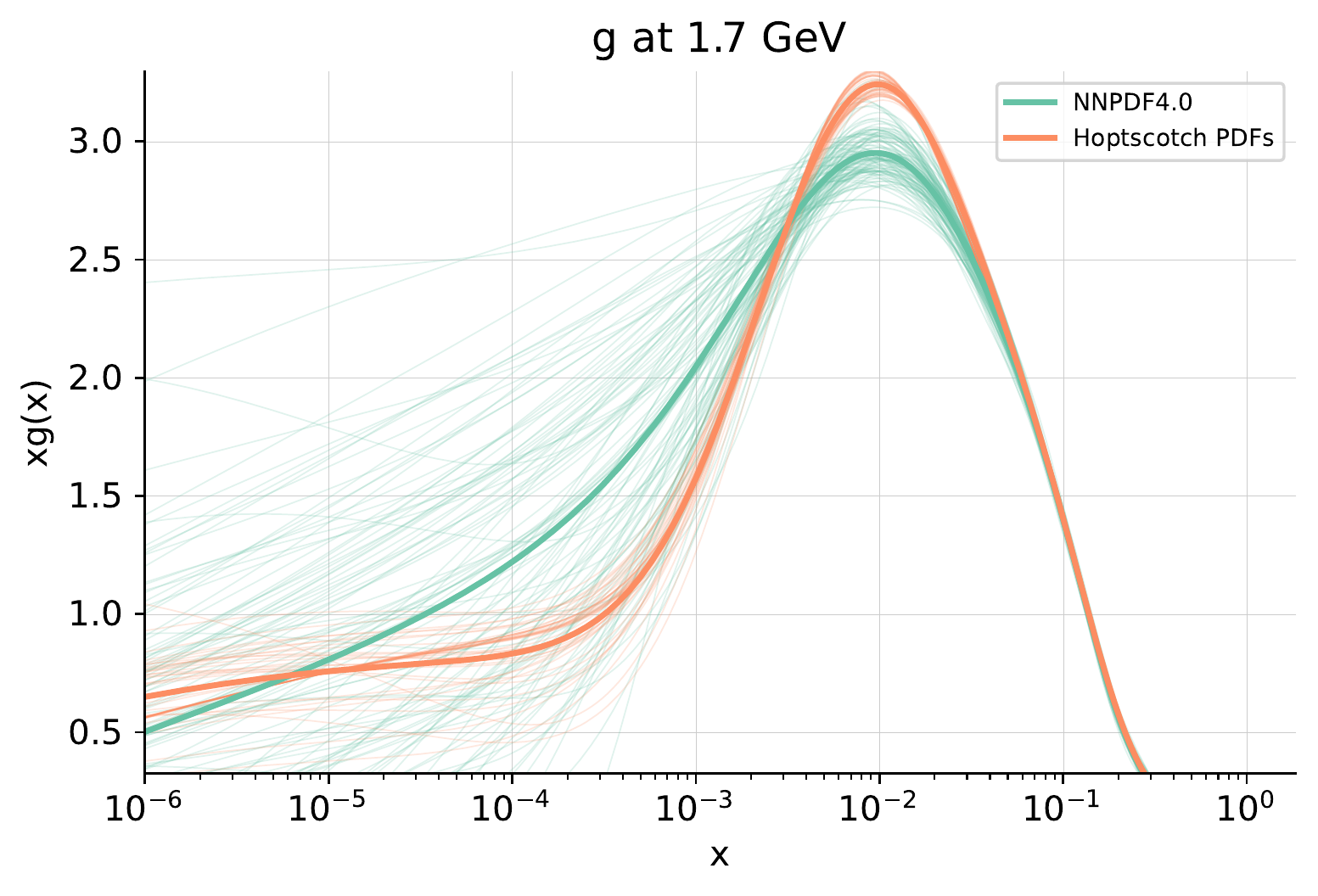}
   \caption{The gluon for HS PDFs with low $\cenchis$ values (orange)
     compared to the NNPDF4.0 gluon (green) 68\%c.l. (left) and 1000
     replica sample
     (right).}
  \label{fig:hsoverfit}
\end{figure}
Having shown that the NNPDF4.0 methodology has no particular difficulty
in fitting the HS PDF, we now
address the question of why these PDFs are unlikely in the NNPDF
methodology. 

\begin{figure}
  \center
  \includegraphics[width=0.49\textwidth]{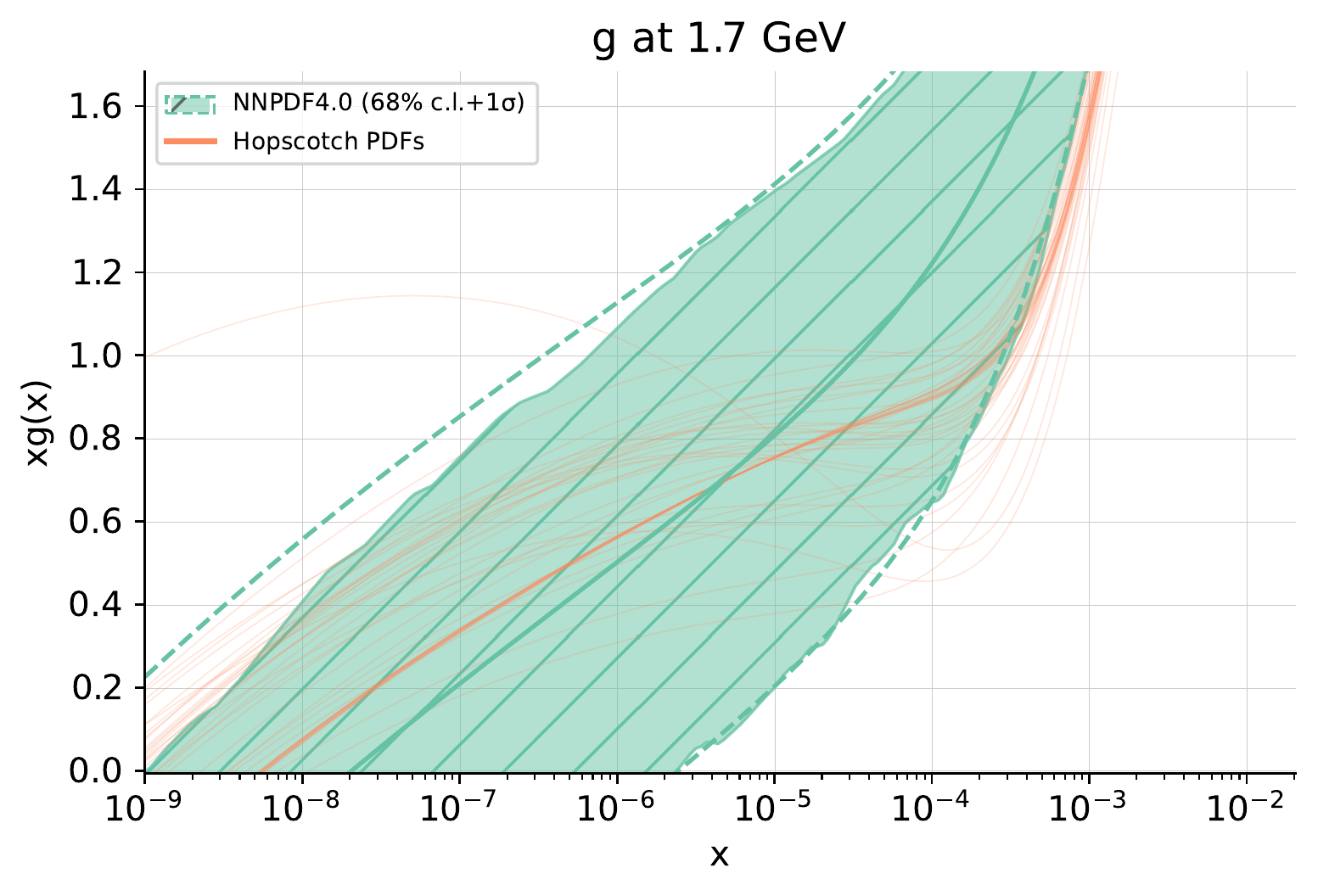}
  \includegraphics[width=0.49\textwidth]{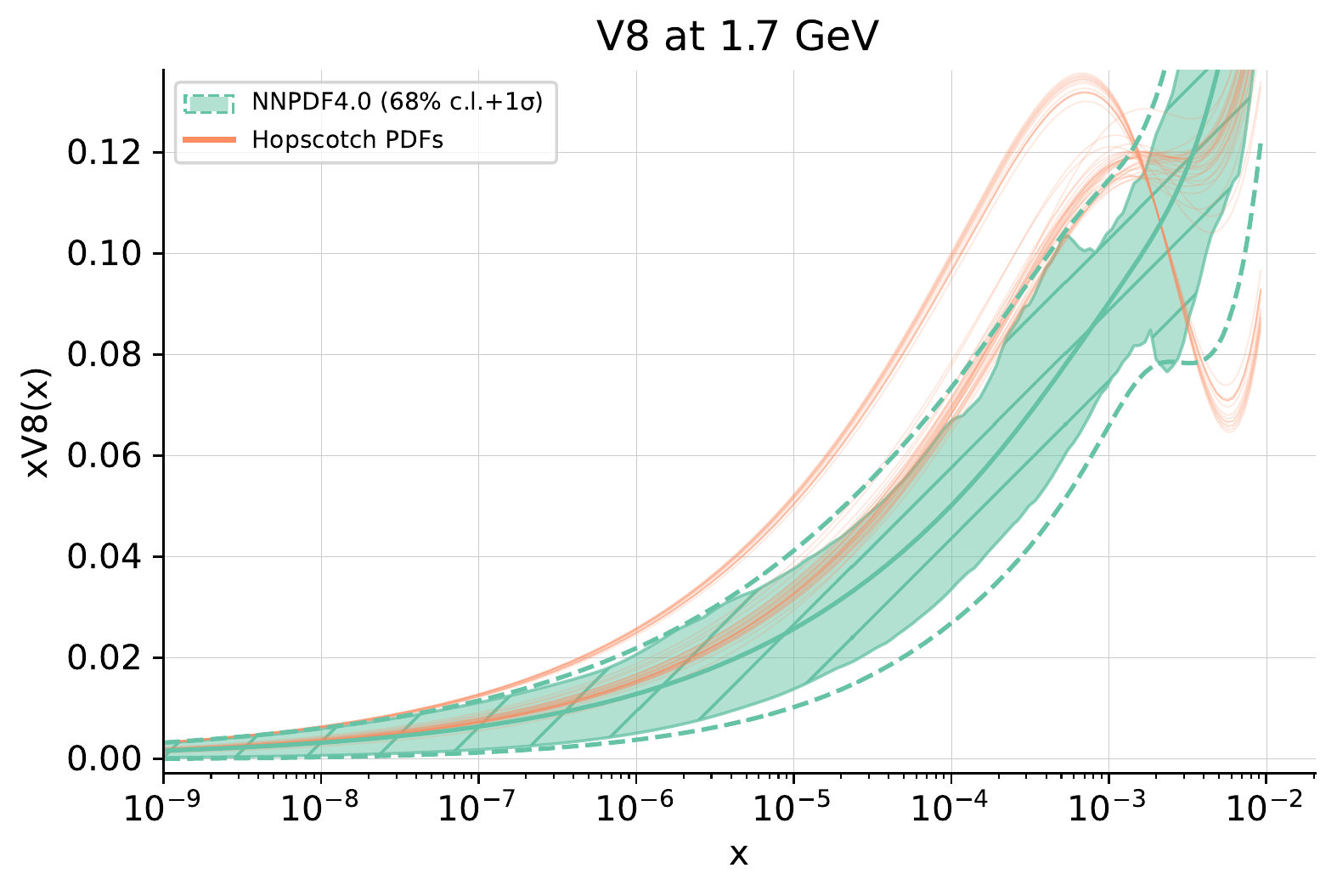}
  \caption{Left: same as the left plot of Fig.~\ref{fig:hsoverfit}, now
    showing a detail of the small-$x$ region. Right: the same as the
    left plot, but now for the valence octet PDF.}
  \label{fig:smallx_gluon}
\end{figure}
As a preliminary observation, we note that, given the extremely
flexible NNPDF parametrization (with about 800 free parameters), it is
very easy to obtain fits with a value of $\cencenchis$ that is much
lower than that of the reference NNPDF4.0 determination, and that
provide an overfit to the data, i.e. fit features of the specific
datasets or processes that do not generalize to other cases. The NNPDF4.0
methodology is carefully tuned in order to avoid such overfits, by a
K-folding procedure~\cite{NNPDF:2021njg} that checks the power of the
PDFs to correctly generalize, and, as already mentioned, tested a posteriori through closure
tests~\cite{NNPDF:2014otw,DelDebbio:2021whr} (that check generalizing power in the data region) and future
tests~\cite{Cruz-Martinez:2021rgy} (that check generalizing power in
the extrapolation region) --- see Appendix~\ref{sec:RegVal} for a
brief review.

As an example of such overfits,  in Fig.~\ref{fig:overfit} we
show the gluon distribution obtained in a fit in which we have
artificially modified the minimization procedure in order to obtain a
very low value of $\cencenchis$. Indeed, in this overfit  the final value
of $\cencenchis$ is by about $\delta=0.08$ smaller than that of the default
NNPDF4.0, i.e. $\delta \times N_{\rm dat}\approx 300$, a difference that is
about one order of magnitude 
bigger than that of the HS PDFs. The unphysical behavior of the PDFs
thus obtained  is manifest and representative of overfitting.

In light of this observation, in Fig.~\ref{fig:hsoverfit} we compare
the NNPDF4.0 PDF gluon PDF to the HS PDFs: for NNPDF we show both the
central value and one-$\sigma$ uncertainty (left plot) and the
corresponding replica set (right), while for HS we can only show the
set of individual PDFs since their ensemble has no statistical
meaning. It is apparent that the HS PDFs are characterized by a kink
in the region $10^{-5}\lesssim x\lesssim 10^{-3}$. This kink is absent
both in the central NNPDF4.0 gluon, and also in individual replicas

In Fig.~\ref{fig:smallx_gluon} we show a detail of the small $x$ region,
in which the kink is clearly visible.
For comparison, in  Fig.~\ref{fig:smallx_gluon} we also show
a similar comparison for the octet valence combination ($V_8=u^-+d^--2
s^-$, with $q_i^-=q-\bar q$) in which even more pronounced kinks are
seen in the HS PDFs.

\begin{figure}
  \center
  \includegraphics[width=0.6\textwidth]{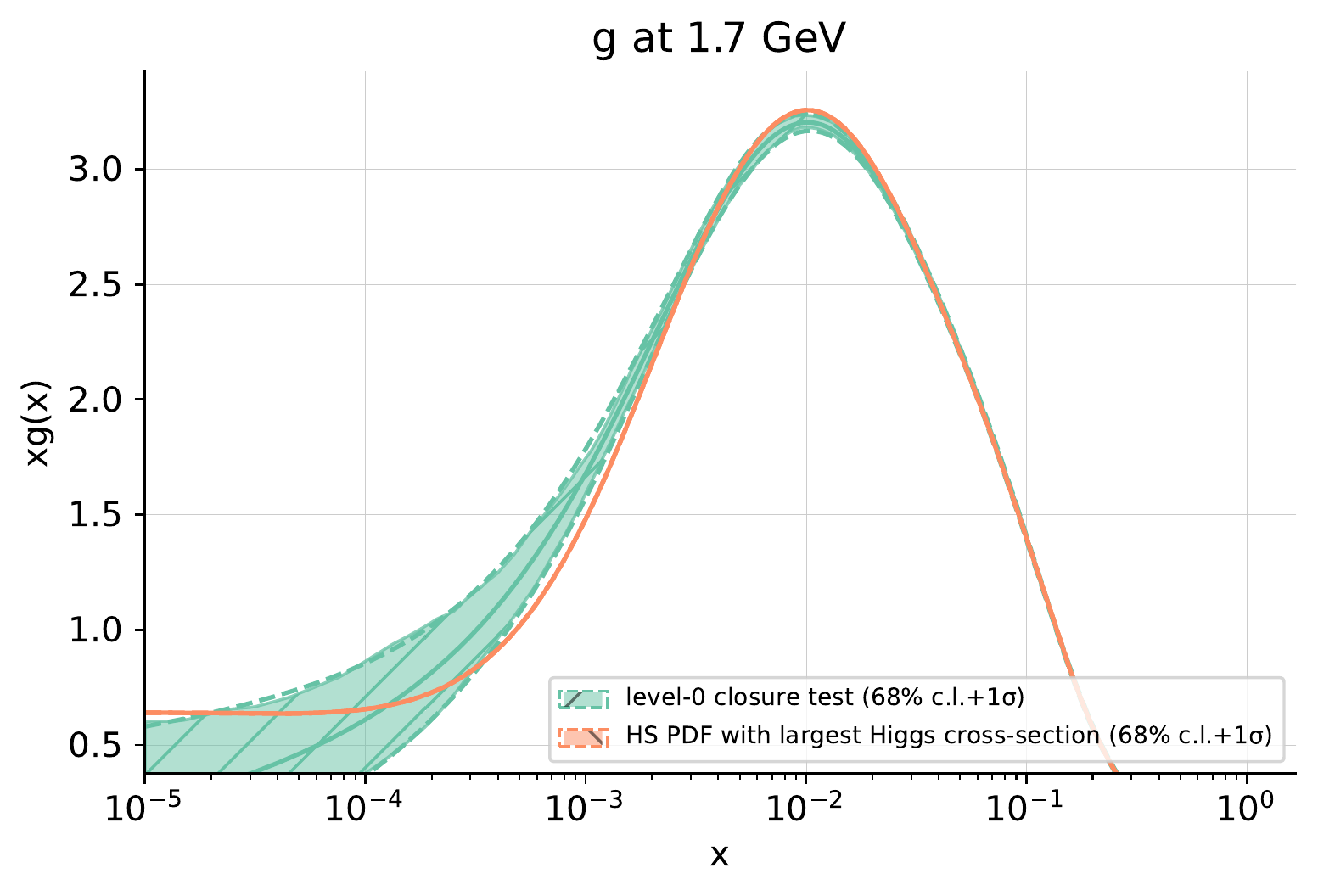}
   \caption{The gluon PDF obtained from the level-0 PDF replicas of
      Fig.~\ref{fig:level0_scatterplot} (green), compared to the underlying
      assumed truth, namely  the
     HS PDF with largest Higgs cross-section taken as underlying
     truth (orange).}
  \label{fig:level0_gluon}
\end{figure}
It is important to observe that there are essentially no data
constraining the gluon in the region of $x\lesssim 10^{-4}$, so the
kink displayed by the HS PDFs is likely not data-driven. We can
actually prove this  explicitly by looking at the PDFs obtained in
the level-0 closure determination of
Fig.~\ref{fig:level0_scatterplot}. Recall that this PDF produces a
perfect fit to the data, i.e. it has vanishing $\chi^2$. In
Fig.~\ref{fig:level0_gluon} the gluon PDF
from this set is shown and compared to the underlying assumed truth HS
PDF. Even though the assumed truth has the kink that characterizes the HS PDFs,
a perfect fit to data as predicted by a HS PDF has no kink. This indicates that
the HS kink is not data driven, but rather an overfitted feature.

\begin{figure}
  \center
  \includegraphics[width=0.49\textwidth]{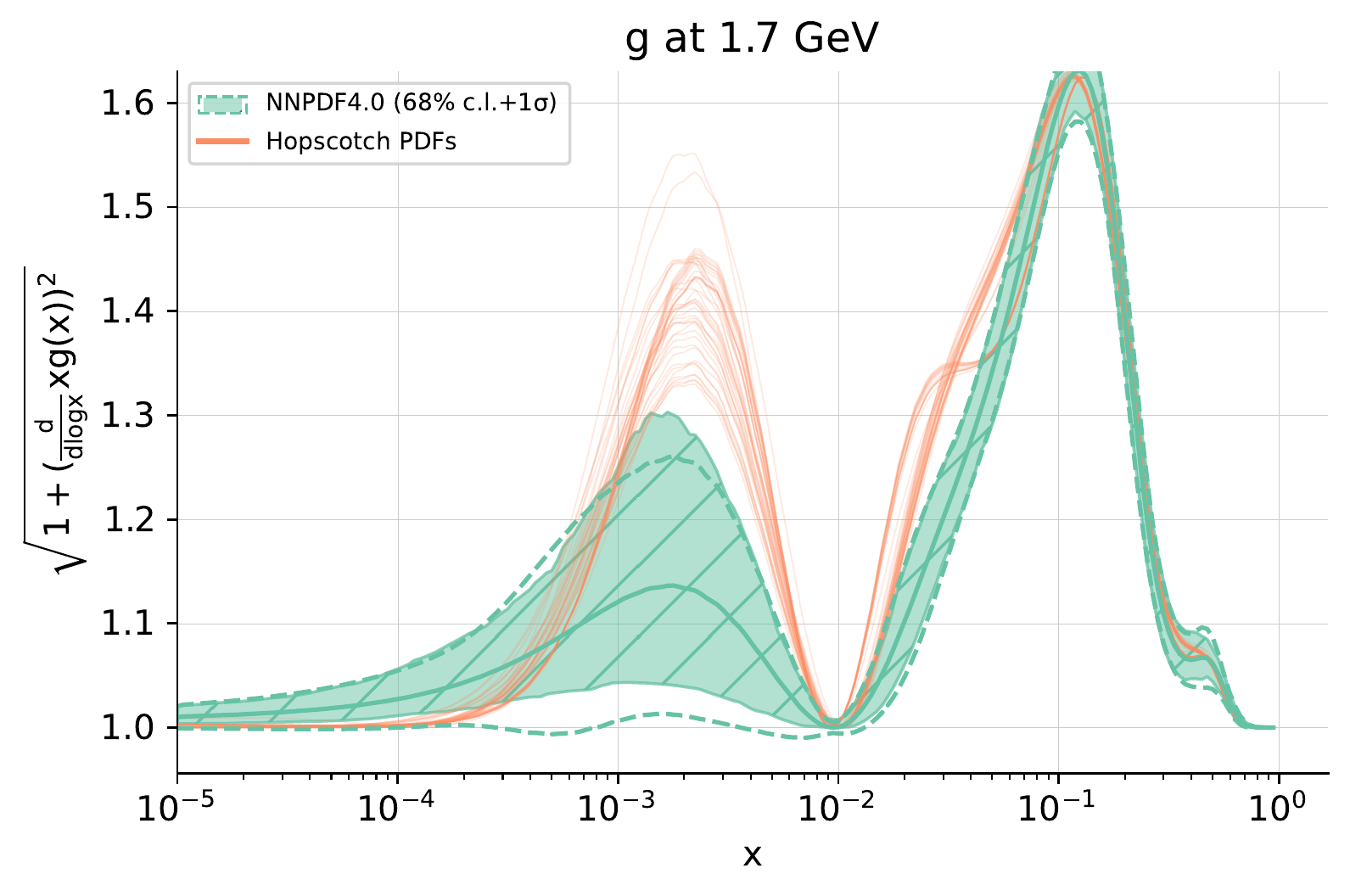}
  \includegraphics[width=0.49\textwidth]{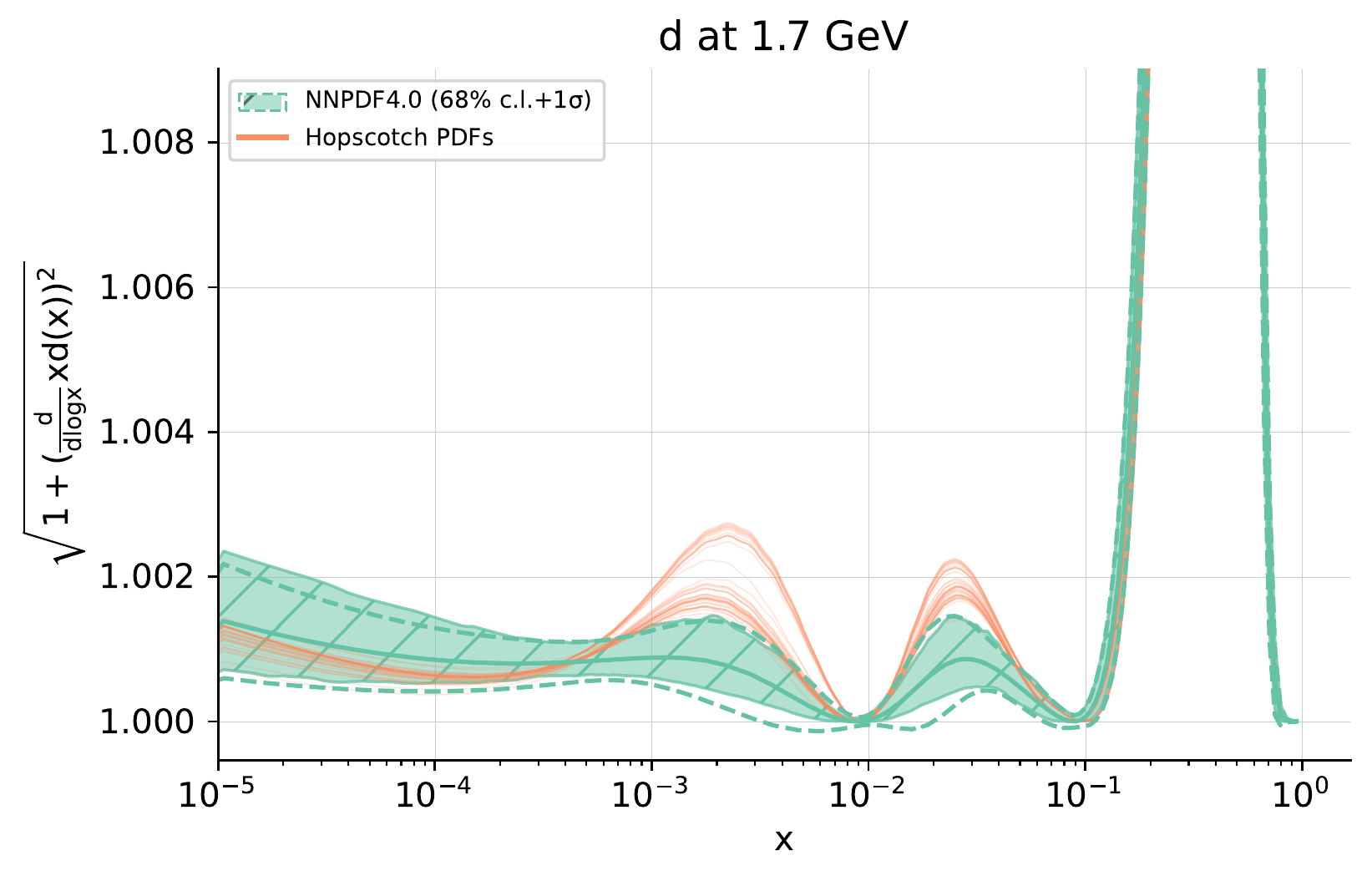}
  \caption{The kinetic energy Eq.~(\ref{eq:arclength}) for the gluon (left)
    and down PDFs (right), for NNPDF4.0 (green) and for the HS PDFs (red).}
  \label{fig:kelog}
\end{figure}
We can actually construct a quantitative overfitting estimator by defining
the PDF kinetic energy
\begin{equation}
{\rm KE}=\sqrt{1+ \left(\frac{d}{d\ln x}xf(x,Q^2)\right)^2}.
\label{eq:arclength}
\end{equation}
This quantity, integrated between any two values of $x$ gives the
arclength of the curve that $x f(x)$ traverses, viewed as a function
of $\ln x$. The kinetic energy is thus a local measure of
``wiggliness'': given a pair of curves with fixed extremes, the one
with greater kinetic energy joins the two extremes with a longer
curve. It coincides of course with the Lagrangian of a relativistic
free particle (with $xf$ interpreted as space and $\ln x$ as time),
hence its name, with its integral being equal to the action.

In Fig.~\ref{fig:kelog} we compare the kinetic energy of the HS PDFs
to that of  NNPDF4.0: we show the gluon and also, for reference, the
down quark. It is clear that the HS PDFs are characterized by higher
kinetic energy, specifically for the gluon, but in fact for all HS
PDFs. Furthermore, the kinetic energy itself displays greater
fluctuations for the HS PDFs. We conclude that the HS PDFs are
characterized by having a feature which is not data driven and that
corresponds to the given curve being further away from a least-action
geodesic, which is disfavoured by the NNPDF methodology. 

The fact that the HS PDFs display signs of overfitting should not come
as a surprise, given that they have been constructed by starting from
NNPDF4.0 replicas, which have been constructed in such a way as to
avoid overfitting, and trying to further minimize the figure of
merit. This suggests that PDF replicas with features similar to the HS
PDFs could be obtained by forcing overfitting in the NNPDF4.0
methodology.

\begin{figure}
  \center
  \includegraphics[width=0.49\textwidth]{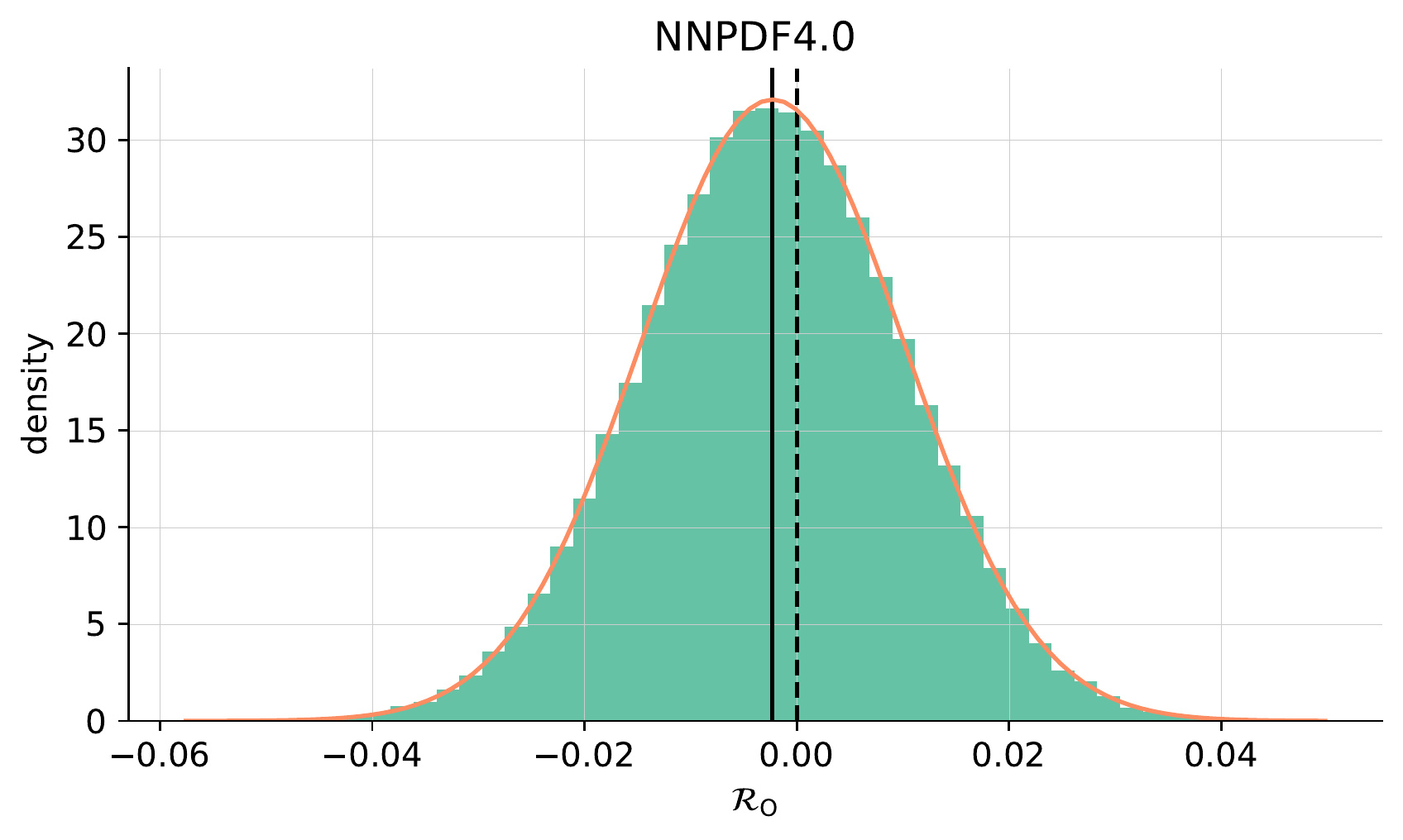}
  \includegraphics[width=0.49\textwidth]{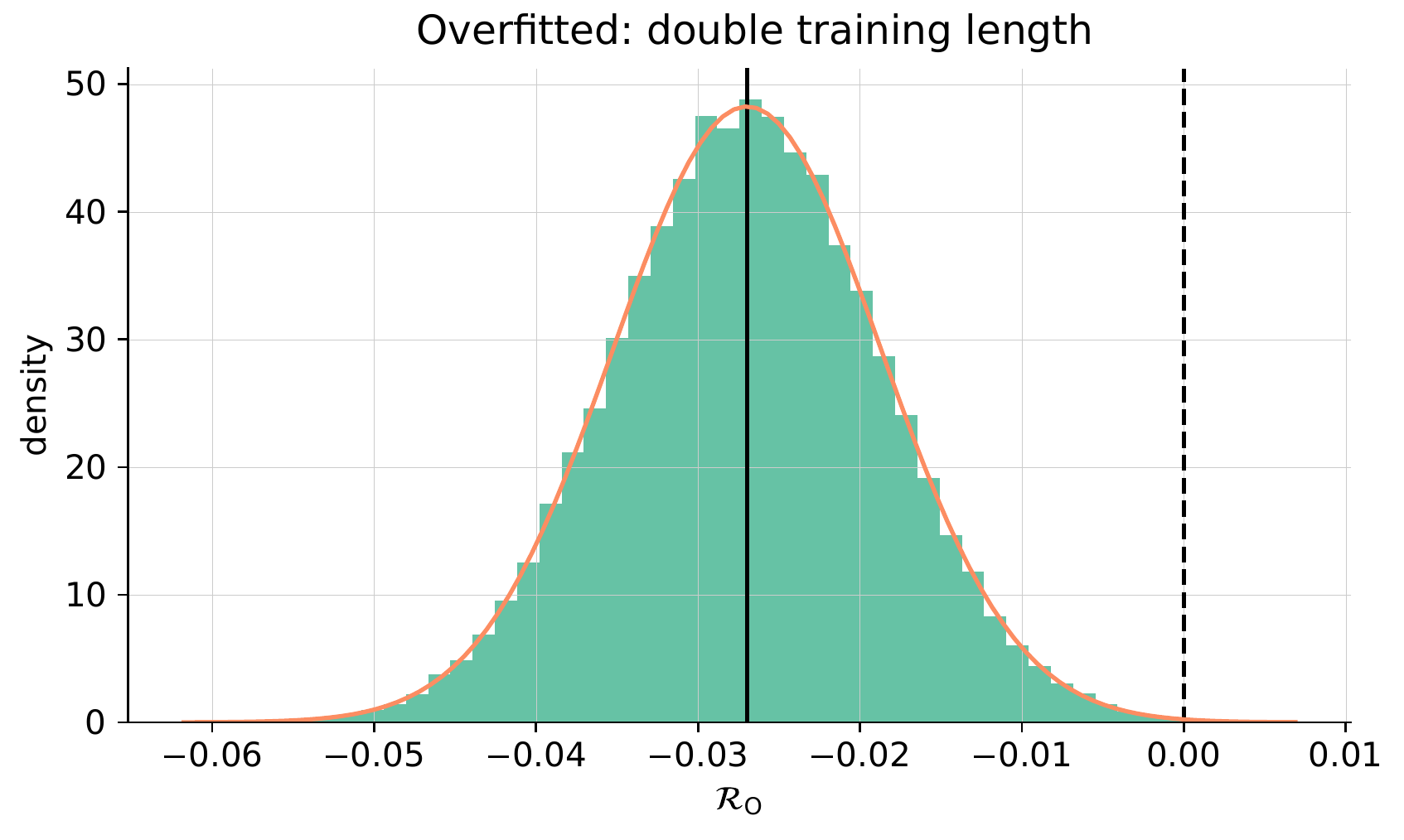}
  \caption{The overfitting metric Ref.~\cite{royness} for the default
    NNPDF4.0 fit and for the artificially overfitted variant discussed
    in text (right).} 
  \label{fig:royness}
\end{figure}
In order to check this explicitly, we have introduced overfitting in
the NNPDF4.0 methodology by changing by hand the fit
settings (which are set by the hyperoptimization procedure). In particular, we doubled the training length, which in
turn implies increasing some parameters (such as the stopping
patience) that are determined as functions of the training length. This
leads to a decrease of $\cencenchis$ by about 0.01, i.e. similar to
the greatest reduction observed in the HS PDFs. We can explicitly check
that these PDFs are overfitted by using an overfitting metric
suggested in Ref.~\cite{royness}, and reviewed in
Appendix~\ref{sec:royness}.
This metric vanishes for a proper
fit, and it is negative for an overfit.
Results are shown in
Fig.~\ref{fig:royness}, where we compare this metric for the default
NNPDF4.0 PDFs and for this overfitted variant. We see that while for
the default $R_{\cal O}=-0.001\pm0.013$, for the overfitted variant 
 $R_{\cal O}=-0.027\pm0.001$, which indicates an overfit at the three
$\sigma$ level.

\begin{figure}
  \center
  \includegraphics[width=0.6\textwidth]{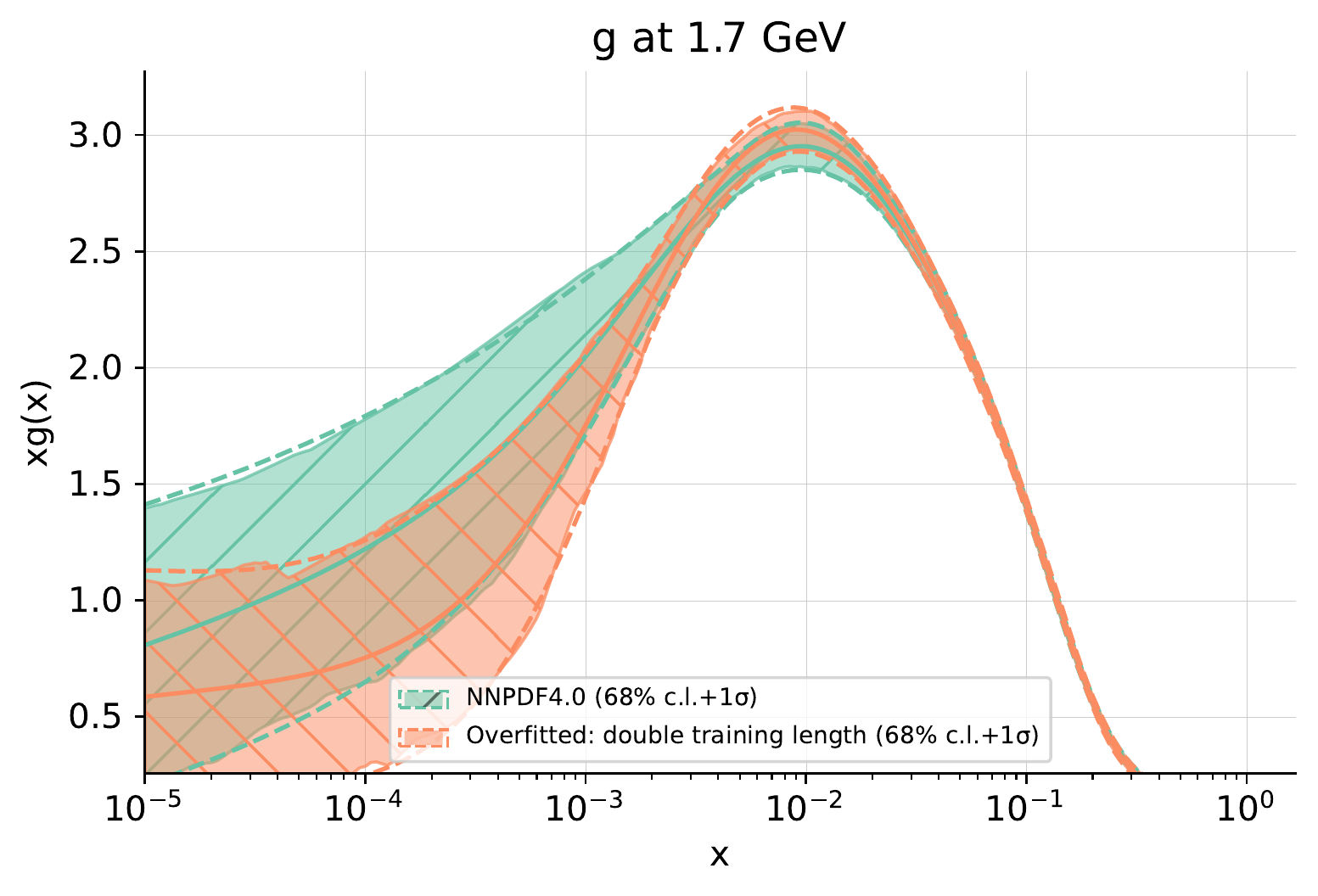}
   \caption{The gluon PDF obtained from the artificially overfitted
     variant discussed 
    in text, compared to the default NNPDF4.0 gluon.}
  \label{fig:overfitg}
\end{figure}
A comparison of the gluon PDF in this overfitted variant to the
default NNPDF4.0 gluon in Fig.~\ref{fig:overfitg} shows that it starts
developing features that are similar to that of the HS PDFs,
specifically a kink in the small $x$ region and a somewhat higher
peak. This provides further evidence that the HS PDFs are overfitted.

Summarizing, we have shown that the HS gluon is  characterized by a feature that
is not data-driven and that corresponds to being further away from a
least-action curve, and that similar features can be obtained in NNPDF4.0
replicas by forcing overfitting. We conclude that NNPDF4.0 replicas that look
like the HS PDFs are disfavored by the NNPDF methodology because they correspond
to overfitting solutions. Nevertheless, PDF replicas leading to results similar
to the HS PDFs in the ZH plane (i.e. leading to similar values of the Higgs and
Z cross section) can be obtained as proper fits to unlikely data fluctuations,
given a large enough replica sample.


\section{Conclusion}

The main purpose of this note was to show that the NNPDF4.0 PDFs are constructed
to give a representative sampling of the probability distribution of PDFs,
constrained by the experimental dataset that enter the fit and the well-known
theoretical constraints given by QCD, such as momentum and valence sum rules, as
well as integrability and positivity. We have explicitly examined the
distribution of the NNPDF4.0 replicas and the fit quality of each replica. The
faithfulness of the NNPDF sampling is confirmed by several tests, for example
the closure tests and future tests, which guarantee that the accuracy of the
sampling is consistent with its stated precision. We have considered an
alternative sampling proposed in Ref.~\cite{Courtoy:2022ocu}. We have explained
that the features of these PDFs can be reproduced by NNPDF4.0 replicas, but with
low posterior probability. We show that such a low probability is related with
the fact that hopscotch PDFs are overfitted, namely they
achieve a low $\chi^2$ to experimental data by fitting the random fluctuations
in the data. The extra wiggliness in the hopscotch PDFs is displayed explicitly,
and explains why they are relatively improbable.
\subsection*{Acknowledgments}

We thank Pavel Nadolsky and Gavin Salam for discussions. We are
grateful to Christopher Schwan for providing the grids used in the
computation of cross-sections.\\
R.~D.~B., L.~D.~D. and partially R.~S. are  supported by the U.K.\
Science and Technology Facility Council (STFC) grant ST/P000630/1.
 S.~F. and partially R.~S. are supported by
the European Research Council under
the European Union's Horizon 2020 research and innovation Programme
(grant agreement n.740006).
Z.~K.~ and M.~U. are supported
by the European Research Council under the European Union’s Horizon 2020 research and
innovation Programme (grant agreement n.950246) and partially supported by the STFC consolidated grant ST/L000385/1; M.~U. is also partially supported by
the Royal Society grant RGF/EA/180148 by the Royal Society grant DH150088.
E.~R.~N.\ is supported by the Italian Ministry of University and Research (MUR)
through the ``Rita Levi-Montalcini'' Program.
J.~R.\ is partially supported by NWO (Dutch Research Council).

\appendix
\section{Validation in NNPDF}
  \label{sec:RegVal}

To validate the resulting PDF distributions, several tests of the NNPDF
methodology have been developed. Specifically, the validity of the PDFs in the
data region is validated using closure tests~\cite{NNPDF:2014otw,DelDebbio:2021whr}  and the validity of the PDFs in the
extrapolation region is validated using futures
tests~\cite{Cruz-Martinez:2021rgy}. We briefly review these tests here, while
referring to the original publications for a more detailed treatment.

\paragraph{Closure test.}
In a closure test, instead of fitting to experimental data, a fit to pseudodata
constructed from predictions using a known input PDF is performed. Closure tests
allow us to validate our methodology by testing if it is able to faithfully
reproduce the underlying input PDF and have been a characteristic of NNPDF fits
since they were first introduced in Ref.~\cite{NNPDF:2014otw}. In order to
quantify the accuracy of the posterior distribution, a number of statistical
estimators are considered that we will discuss here. These statistical
estimators have been evaluated for the NNPDF4.0 determination, and found to be
within 1$\sigma$ of the value corresponding to a faithful representation of the
input PDF. For a detailed motivation of these estimators we refer the reader to
Ref.~\cite{NNPDF:2021njg}.

To estimate the faithfulness of the PDF uncertainty at the level of observables,
the bias over variance ratio as defined in Eq.~(6.15) of
Ref.~\cite{NNPDF:2021njg} is used. Here ``bias'' can be understood as a measure
of the fluctuations of the observable values with respect to the central value
prediction of the fitted PDF, while ``variance'' can be understood as the
fluctuations of the fitted PDF with respect to its central value prediction.
Thus if the methodology has faithfully reproduced the uncertainties in the
underlying data (bias), this uncertainty should be equal to the uncertainty in
the predictions of the PDFs (variance), and hence the bias to variance ratio
$R_{\rm bv}$ is expected to be one.


To estimate the faithfulness of the PDF uncertainty at the level of the PDF we
calculate a quantile estimator in PDF space $\xi^{(\rm pdf)}_{1\sigma}$. This
quantity corresponds to the number of fits for which the $1\sigma$ uncertainty
band covers the PDF used as underlying law. The expected value is 0.68 for $\xi^{(\rm pdf)}_{1\sigma}$, 0.95 for  $\xi^{(\rm pdf)}_{2\sigma}$, etc. 

An analogous estimator can be calculated for the theory predictions in data
space as opposed to PDF space, providing a generalization to quantile statics of
the bias of variance ratio $R_{\rm bv}$. 
The expected value of this quantile estimator depends on the bias over variance
ratio and is $\mathrm{erf} (R_{\rm bv}/\sqrt{2})$, which is checked to be in
agreement with the calculated value of $\xi^{\rm (exp)}_{1\sigma}$.


Finally it should be noted that for each of these estimators the corresponding
standard error on the calculated value is determined through a bootstrapping
procedure~\cite{Efron:1979bxm,Efron:1986hys}

\paragraph{Future test.}
By definition, testing the accuracy in a region where there is no data to test
against is impossible. Thus to test the accuracy of the predictions in the
extrapolation region, a fit is performed using a ``historic'' dataset
representing the knowledge available at an earlier point in time and its
predictions compared to the more recent measurements. This is the basic premise
of the the ``future test'' technique introduced in
Ref.~\cite{Cruz-Martinez:2021rgy}. Specifically, the historic datasets for the
validation performed for the NNPDF4.0 determination correspond to a pre-HERA and
pre-LHC period where. For a detailed overview of exactly which datasets are
included in each of the historic datasets we refer to Tab.~6.5 of
Ref.~\cite{NNPDF:2021njg}.

Since the aim of doing a future test is to determine the ability of a
methodology for PDF determination to provide a generalized fit, we need to take
into account not only the uncertainty of the experimental data but also the
uncertainty of the PDF itself. This is done by adding to the experimental
covariance matrix (as will be defined in Sect.~\ref{sec:misleading}) also the
covariance matrix of the observables calculated from PDF predictions:
\begin{equation}
  \left(\operatorname{cov}_{\textrm{future test}}\right)_{i j} = \left(\operatorname{cov}_{\rm exp}\right)_{i j}  + \left(\operatorname{cov}_{\rm pdf }\right)_{i j},
    \label{eq:cov=exp+pdf}
\end{equation}
Where $\left(\operatorname{cov}_{\rm pdf }\right)_{i j}$ is defined as
\begin{equation}
  \left(\operatorname{cov}_{\rm pdf }\right)_{i j} = \langle \mathcal{T}_i\mathcal{T}_j  \rangle_{\rm rep} - \langle \mathcal{T}_i  \rangle_{\rm rep}\langle \mathcal{T}_j  \rangle_{\rm rep},
\end{equation}
with $\mathcal{T}^{(k)}_i$ the prediction of the $i$-th datapoint using the
$k$-th PDF replica with the average defined over replicas.

While in general the expected disagreement between the predication of the unseen
(or future) data and the actual data will be significant, the future test allows
us to test whether methodology properly accounts for this by correspondingly
increasing the PDF uncertainties. The $\chi^2$ as defined using the future test
covariance matrix constructed by combining the PDF and experimental covariance
matrices per Eq.~(\ref{eq:cov=exp+pdf}) should reduce to the usual value of
order one.

\section{An overfitting metric}
\label{sec:royness}

If the optimal fit is determined through a
cross-validation stopping criterion, based on splicing the data in
training and validation sets, it is possible to construct a metric
that explicitly detects the presence of overfitting in the final result
~\cite{royness}. We review here the way this metric is constructed.

Recall that the training-validation split is done randomly on a
replica-by-replica basis. This means that for each replica, different
data go into the training and validation sets. We call the way the
data are split into training and validation sets a ``validation
mask''. So each data replica is characterized by the  fact that (a)
data are fluctuated differently and (b) a different mask is adopted.
Based on this observation, define
\begin{equation}
  \overline{\chi_\mathrm{val}^2}\left[\mathcal{T}^{(k)},
    F^{(k)}\right] \equiv
  \frac{1}{N}\sum_{k'=1}^{N}\chi_\mathrm{val}^2\left[\mathcal{T}^{(k)},
    F^{(k')}\right] \quad\Bigg|_{\rm fixed\,mask}.
  \label{eq:overfitting_equality}
\end{equation}
The sum here is performed over different data replicas, but with a
fixed validation mask: i.e. the data are fluctuated differently, but
the same validation mask is adopted.

Now define
\begin{equation}
  \mathcal{R}_O=\chi_\mathrm{val}^2\left[\mathcal{T}^{(k)}, F^{(k)}\right] -  \overline{\chi_\mathrm{val}^2}\left[\mathcal{T}^{(k)},
    F^{(k)}\right],
  \label{eq:overfitting_metric}
\end{equation}
where $\chi_\mathrm{val}^2\left[\mathcal{T}^{(k)}, F^{(k)}\right]$
is the usual validation $\chi^2$ for the fitting of the $k$-th
replica, and  $\overline{\chi_\mathrm{val}^2}\left[\mathcal{T}^{(k)},
    F^{(k)}\right]$ is computed using (for all replicas) the same
validation mask that was used for this $k$-th replica.

If a PDF replica $f^{(k)}$ is overfitted, i.e. it does contain some
information specific of the underlying data replica $F^{(k)}$, then
the quantity $\mathcal{R}_O$ is negative, meaning that its $\chi^2$ is
lower than that which is obtained on average when comparing to  a
random fluctuation of the same validation data. If instead it is
correctly fitted, then $\mathcal{R}_O$ should be compatible with zero.

\bibliographystyle{JHEP}
\bibliography{response}
\end{document}